\documentclass[a4paper,onecolumn,unpublished]{quantumarticle}
\pdfoutput=1
\PassOptionsToPackage{hyphens}{url}
\usepackage[usenames,dvipsnames]{xcolor}
\usepackage{graphicx}
\usepackage[ colorlinks = true, 
             linkcolor = Periwinkle,
             urlcolor  = ForestGreen!85!black,
             citecolor = ForestGreen!85!black,
             anchorcolor = ForestGreen!85!black,
]{hyperref} 

\usepackage{amsmath,amssymb,amsthm,mathtools,mathrsfs}
\usepackage{pifont}
\usepackage{authblk} 
\usepackage{bbm}
\usepackage{xparse} 
\usepackage{physics}
\usepackage{tikz}
\usepackage[compat=1.1.0]{tikz-feynman}
\usepackage{dsfont}
\usepackage{enumerate}
\usepackage{tocbasic}

\DeclareTOCStyleEntry[
  beforeskip=0.3em plus 1pt,
  pagenumberformat=\textbf
]{tocline}{section}

\makeatletter
\renewcommand*\env@matrix[1][\arraystretch]{%
  \edef\arraystretch{#1}%
  \hskip -\arraycolsep
  \let\@ifnextchar\new@ifnextchar
  \array{*\c@MaxMatrixCols c}}
\makeatother

\theoremstyle{plain}
\newtheorem{theorem}[equation]{Theorem}
\newtheorem{lemma}[equation]{Lemma}
\newtheorem{proposition}[equation]{Proposition}
\newtheorem{corollary}[equation]{Corollary}
\theoremstyle{definition}
\newtheorem{definition}[equation]{Definition}

\newtheorem{question}[equation]{Question}

\newtheorem{problem}[equation]{Problem}
\newtheorem{example}[equation]{Example}
\newtheorem{exercise}[equation]{Exercise}
\newtheorem*{answer}{Answer}
\newtheorem*{solution}{Solution}
\newtheorem{remark}[equation]{Remark}
\newtheorem{terminology}[equation]{Terminology}

\newtheorem{notation}[equation]{Notation}
\newtheorem{noterm}[equation]{Notation and Terminology}

\newcommand\define[1]{\emph{\textbf{#1}}}

\numberwithin{equation}{section}

\newcommand{\be}{\begin{equation}}
\newcommand{\ee}{\end{equation}}
\newcommand{\bx}{\begin{example}}
\newcommand{\ex}{\end{example}}
\newcommand{\bex}{\begin{exercise}}
\newcommand{\eex}{\end{exercise}}
\newcommand{\ban}{\begin{answer}}
\newcommand{\ean}{\end{answer}}
\newcommand{\bt}{\begin{theorem}}
\newcommand{\et}{\end{theorem}}
\newcommand{\bc}{\begin{corollary}}
\newcommand{\ec}{\end{corollary}}
\newcommand{\blem}{\begin{lemma}}
\newcommand{\elem}{\end{lemma}}
\newcommand{\bp}{\begin{problem}}
\newcommand{\ep}{\end{problem}}
\newcommand{\bn}{\begin{proposition}}
\newcommand{\en}{\end{proposition}}
\newcommand{\bd}{\begin{definition}}
\newcommand{\ed}{\end{definition}}
\newcommand{\bq}{\begin{question}}
\newcommand{\eq}{\end{question}}
\newcommand{\bprf}{\begin{proof}}
\newcommand{\eprf}{\end{proof}}
\newcommand{\br}{\begin{remark}}
\newcommand{\er}{\end{remark}}
\newcommand{\bs}{\begin{solution}}
\newcommand{\es}{\end{solution}}
\newcommand{\bnt}{\begin{noterm}}
\newcommand{\ent}{\end{noterm}}
\newcommand{\bnot}{\begin{notation}}
\newcommand{\enot}{\end{notation}}
\newcommand{\bterm}{\begin{terminology}}
\newcommand{\eterm}{\end{terminology}}

\newcommand{\id}{\mathrm{id}}

\def\R{{{\mathbb R}}}
\def\C{{{\mathbb C}}}

\def\Q{{{\mathbb Q}}}

\newcommand{\Ad}{\mathrm{Ad}}

\DeclareMathAlphabet{\mathpzc}{OT1}{pzc}{m}{it} 
 \DeclareFontFamily{OT1}{pzc}{}
 \DeclareFontShape{OT1}{pzc}{m}{it}{ <-> s*[1.2] pzcmi7t }{}
 \DeclareMathAlphabet{\mathpzc}{OT1}{pzc}{m}{it}

\newcommand{\Lin}{\mathcal L}
\newcommand{\one}{\mathds 1}

\def\E{\mathcal{E}}
\def\H{\mathcal{H}}
\def\F{\mathcal{F}}
\def\O{\mathscr{O}}

\def\J{\mathscr{J}}

\def\C{\mathbb{C}}
\def\Q{\mathcal{Q}}
\def\R{\mathbb{R}}

\title{On spacetime quantum states}
\author{James Fullwood}
\affiliation{School of Mathematics and Statistics, Hainan University, Haikou, Hainan Province, 570228, China}
\affiliation{Hainan International Exchange Center for Theoretical Physics, Haikou, Hainan Province, 570228, China}
\affiliation{Key Laboratory of Engineering Modeling and Statistical Computation of Hainan Province, Hainan University, Haikou 570228, China}
\email{fullwood@hainanu.edu.cn}

\begin{document}
\emergencystretch 2em

\maketitle

\vspace{-7mm}
\tableofcontents

\begin{abstract}
Recent efforts to extend the density operator formalism into the temporal domain have led to several non-Hermitian proposals across quantum information theory, quantum foundations, and quantum field theory. In this work, we prove that three such constructions, namely, Kirkwood-Dirac (KD) states over time, spacetime states, and spacetime density matrices, are all identical up to taking Hermitian adjoints. For time-homogeneous unitary evolution, we show that spacetime states coincide with KD states over time across any number of time slices, while for quantum dynamics given by a sequence of quantum channels we prove that the multi-time spacetime density matrix is precisely the Hermitian adjoint of the corresponding KD state over time. Together, these equivalences establish a direct dictionary between independently developed frameworks, while showing that KD states over time provide a natural generalization of both spacetime states and spacetime density matrices to general open-system dynamics. In light of these results, the terms \emph{spacetime state} and \emph{spacetime density matrix} may be used interchangeably up to Hermitian conjugation, just as \emph{quantum state} and \emph{density matrix} are for standard spatial quantum states. Building on the fact that every such spacetime state is a unit-trace contraction, we also establish an axiomatic derivation of \emph{spacetime entropy} for spacetime states, extending both von~Neumann entropy and pseudo-density matrix entropy to the non-Hermitian domain. Specifically, we prove that four conditions---reduction to von~Neumann entropy, continuity, strong additivity, and the \emph{polar expectation value} property---isolate a unique entropy functional on finite-dimensional unit-trace contractions. It is also shown that this spacetime entropy avoids the branch-cut ambiguities of constructions based on analytic continuation of the logarithm (such as pseudo-entropy), satisfies additivity over tensor products, reduces to the von~Neumann entropy of the initial state under unitary dynamics, and undergoes complex conjugation upon taking Hermitian adjoints.
\end{abstract}

\section{Introduction}

The notion of a `quantum state' is fundamental to quantum theory, dating back to the work of Bohr and Sommerfeld on atomic orbits~\cite{Bohr_1913,Sommerfeld_1916}. In its modern formulation, a general state of a quantum system is represented mathematically by a density operator $\rho$ on a Hilbert space $\H$, from which all physically meaningful statistical data about the system may be extracted. In particular, expectation values, probabilities, and equal-time correlation functions are all encoded in the quantum state of a system. Moreover, with the rise of quantum information theory, information-theoretic quantities associated with a density operator such as entanglement entropy have also come to play a prominent role in our understanding of quantum systems.

Crucially, however, a composite system $AB$ is assigned a joint quantum state only when $A$ and $B$ may be thought of as coexisting at a single instant in time. For example, if $A$ and $B$ represent the same system at two times $t_A$ and $t_B > t_A$, then standard quantum theory does not associate a joint quantum state with the composite system $AB$. While at first glance this may not appear to be a defect of quantum theory, the absence of a joint state across time leads to fundamental limitations, especially in an information-theoretic context. In particular, it prevents a straightforward definition of entanglement across time, confining quantities such as entanglement entropy and quantum mutual information to spacelike separated systems. In this sense, there are many respects in which quantum information theory remains a static theory of information, in stark contrast with classical information theory, which makes no distinction between spacelike and timelike separated random variables.

In response to this asymmetry between space and time in quantum theory, there have been various extensions of the density operator formalism into the temporal domain. While early proposals focused largely on Hermitian operators of unit trace whose single-time marginals recover ordinary density operators---such as pseudo-density matrices~\cite{FJV15} and Hermitian quantum states over time~\cite{LeSp13,HHPBS17,FuPa22}---more recently \emph{non-Hermitian} extensions have been proposed by researchers in quantum information theory, quantum foundations, high-energy physics, and quantum field theory. Within quantum information and foundations, these appear as \emph{Kirkwood-Dirac (KD) states over time} (also termed `left-product states over time'~\cite{Lie_2025,Lie_2026}, `right blooms'~\cite{FuPa22a,Fu_2025}, and `Kirkwood-Dirac temporal states'~\cite{JMK_2026,Jia_Xiao_2026}), which may be constructed directly from the data of an initial state together with a sequence of quantum channels responsible for its evolution. Within the framework of spacetime quantum mechanics~\cite{Diaz_2024}, Diaz et al. define $N$-time \emph{spacetime states} via a quantum action built from cyclic shift operators and the Hamiltonian governing unitary dynamics~\cite{Diaz_2026a,Diaz_2026}. Meanwhile, motivated by timelike entanglement~\cite{DHMTT22,Doi_2023} and relativistic quantum field theory, Milekhin, Adamska, and Preskill~\cite{Milekhin_2025} and Guo~\cite{Guo_2025,Guo_2026} introduce \emph{spacetime density matrices}, defined implicitly as the unique operators encoding $N$-point anti-time-ordered Wightman correlation functions~\cite{Weinberg_1995}.

In this work, we prove that KD states over time, spacetime states, and spacetime density matrices are all mathematically equivalent up to Hermitian conjugation, and we leverage this common framework to establish an axiomatic derivation of spacetime entropy. Given the markedly different motivations, notations, and mathematical machinery underlying these proposals, their equivalence is quite remarkable, foreshadowing a canonical operator formalism for spatiotemporal quantum correlations common to quantum information, quantum foundations, and high-energy physics. Indeed, the independent emergence of the same operator across such disparate lines of inquiry suggests that these non-Hermitian constructions reflect an intrinsic structural feature of quantum dynamics rather than an artifact of any particular framework. Establishing their equivalence also provides a direct dictionary between these approaches, allowing operational insights and open-system methods from quantum information theory to inform studies of spacetime quantum mechanics and timelike entanglement in quantum field theory, and vice versa. Furthermore, viewing these objects through the lens of KD states over time yields a simple closed-form expression that extends directly from unitary evolution to arbitrary open-system dynamics, while revealing that every spacetime state is a unit-trace contraction. This bound on the operator norm in turn provides a natural starting point for extending the axiomatic characterization of pseudo-density matrix entropy from Ref.~\cite{Fullwood_Yang_2026} to the non-Hermitian setting, isolating a unique spacetime entropy functional through four simple axioms and enabling a systematic derivation of its fundamental properties.

In what follows, we make these equivalences precise, establish the basic mathematical structure of the underlying operators, and derive the corresponding spacetime entropy functional together with its core properties. In Section~\ref{sec:kd}, we fix notation and give a self-contained introduction to $N$-time KD states over time $T_N$, reviewing their appearances in the literature as well as their relationship to pseudo-density matrices. We also prove several fundamental properties of $T_N$ in Proposition~\ref{prop:kd-properties} (some of which are well known), namely, the Kirkwood-Dirac correlator identity, unit trace and marginal consistency, convex linearity in the initial state and multilinearity in the channels, local unitary covariance, factorization under parallel composition, and the operator-norm contraction bound $\|T_N\|_\infty \le 1$. In Section~\ref{sec:equiv}, we review the $N$-time spacetime states $R_N$ of Diaz et al.~\cite{Diaz_2026a,Diaz_2026} and prove Theorem~\ref{prop:N-slices}, establishing the Jamio{\l}kowski factorization of the $N$-time quantum action from Ref.~\cite{Diaz_2026} and the resulting equivalence $R_N = T_N$ for time-homogeneous unitary evolution, thereby extending the $N = 2$ identification noted in~\cite{Diaz_2026} to arbitrary $N > 1$. In Section~\ref{sec:sdm-equiv}, we introduce the $N$-time spacetime density matrix $\Sigma_N$ extending the constructions of Milekhin, Adamska, and Preskill~\cite{Milekhin_2025} and Guo~\cite{Guo_2025}, and prove Theorem~\ref{thm:sdm-equiv}, showing that for any $N$-time quantum dynamics one has $\Sigma_N = T_N^\dagger$ (and hence $\Sigma_N = R_N^\dagger$ in the time-homogeneous unitary setting). Taken together, Theorems~\ref{prop:N-slices} and~\ref{thm:sdm-equiv} place the two-time channel construction of~\cite{Milekhin_2025}, the multi-time unitary construction of~\cite{Guo_2025}, and the spacetime states of~\cite{Diaz_2026} within a single framework, while showing that KD states over time and their Hermitian adjoints provide a straightforward generalization of both spacetime states and multi-time spacetime density matrices to arbitrary open-system dynamics. Finally, in Section~\ref{sec:entropy}, we introduce the notion of a \emph{polar expectation value} (Definition~\ref{def:mod-exp}) and prove Theorem~\ref{thm:spacetime-entropy}, showing that reduction to von~Neumann entropy, continuity, strong additivity over the closed unit disk $\mathbb{D}$, and the polar expectation value property isolate a unique entropy functional on the space of finite-dimensional unit-trace contractions, which we refer to as \emph{spacetime entropy}. We then establish several fundamental properties of this spacetime entropy in Proposition~\ref{prop:spacetime-entropy-properties}, namely, additivity over tensor products and parallel dynamics, a singular-value upper bound on $|S(X)|$, exact invariance $S(T_N) = S(\Sigma_N) = S(\rho_1)$ under arbitrary $N$-time isometric and unitary dynamics, and symmetries under Hermitian conjugation, transposition, and entrywise complex conjugation. In particular, spacetime entropy satisfies $S(\Sigma_N) = \overline{S(T_N)}$, so that reversing the operator ordering between $T_N$ and $\Sigma_N$ preserves the real part of the entropy while reversing the sign of its imaginary part.

\section{Kirkwood-Dirac states over time}\label{sec:kd}

In this section we fix notation and give a self-contained introduction to Kirkwood-Dirac (KD) states over time. Throughout this work, let $N > 1$, and let $\H_{1\cdots N} = \H_1 \otimes \cdots \otimes \H_{N}$, where each factor $\H_k$ is a finite-dimensional Hilbert space. The algebra of linear operators on $\H_k$ will be denoted by $\mathcal{L}(\H_k)$, and we let $\one_k \in \Lin(\H_k)$ denote the identity operator on $\H_k$. A linear map $\E:\mathcal{L}(\H_k)\to \mathcal{L}(\H_{k+1})$ which is completely positive and trace-preserving (CPTP) will be referred to as a \emph{quantum channel}.

\bd\label{def:lpsot}
The \define{Jamio{\l}kowski operator} of a linear map $\E \colon \Lin(\H_k) \to \Lin(\H_{k+1})$ is the operator $\J[\E] \in \Lin(\H_k \otimes \H_{k+1})$ given by
\be
\J[\E] = \sum_{i,j} \dyad{i}{j} \otimes \E(\dyad{j}{i})\, ,
\ee
where $\{\ket{i}\}$ is an orthonormal basis of $\H_k$.
\ed

\br
The Jamio\l kowski operator $\J[\E]$ first appears in \cite{Ja72}, and is the partial transpose of the Choi operator $\mathscr{C}[\E]$ with respect to the orthonormal basis $\{\ket{i}\}$ of $\H_k$~\cite{Ch75}. While the Choi matrix is basis-dependent, the Jamio\l kowski operator $\J[\E]$ is in fact a basis-independent construction, as it may be written as $\J[\E]=(\id\otimes \E)({\tt SWAP})$, where ${\tt SWAP}:\H_k\otimes \H_{k}\to \H_{k}\otimes \H_k$ is the swap isomorphism.
\er

\bd\label{def:kd}
Given a density operator $\rho_1 \in \Lin(\H_1)$ and a sequence $(\E_k \colon \Lin(\H_{k}) \to \Lin(\H_{k+1}))_{k=1}^{N-1}$ of quantum channels, the $N$-tuple $(\rho_1,\E_1,\ldots, \E_{N-1})$ is said to be an \define{$N$-time quantum dynamics}. The \define{Kirkwood-Dirac (KD) state over time} associated with an $N$-time quantum dynamics $(\rho_1,\E_1,\ldots, \E_{N-1})$ is the operator $T_{N}\in \Lin(\H_{1\cdots N})$ given by
\be
T_{N} = (\rho_1 \otimes \one_2 \otimes \cdots \otimes \one_{N}) \prod_{k=1}^{N-1} \J_{k,k+1}[\E_k]\, ,
\ee
where $\J_{k,k+1}[\E_k] \in \Lin(\H_{1\cdots N})$ acts as $\J[\E_k]$ on $\H_{k}\otimes \H_{k+1}$ and as the identity on all other factors. 
\ed

\bx
For a $2$-time quantum dynamics $(\rho_1,\E_1)$ we have
\[
T_{2} = (\rho_1 \otimes \one_2)\J[\E_1]\, ,
\]
and for a $3$-time quantum dynamics $(\rho_1,\E_1,\E_2)$ we have
\[
T_{3} = (\rho_1 \otimes \one_2 \otimes \one_3)(\J[\E_1] \otimes \one_3)(\one_1 \otimes \J[\E_2])\, .
\]
\ex

\br[Appearance of KD states over time in the literature]
To the best of our knowledge, the earliest appearance of the KD state over time is in the context of 2-times in Ref.~\cite{LeSp13}, where it was suggested as a non-Hermitian alternative for overcoming the limitations of the Hermitian state over time considered therein. In Ref.~\cite{HHPBS17}, it was shown that the only constructions associated with a 2-time quantum dynamics $(\rho,\E)$ satisfying a certain list of physically motivated axioms were $(\rho\otimes \mathds{1})\J[\E]$ and $\J[\E](\rho\otimes \mathds{1})$, which are the 2-time KD state over time $T_2$ and its adjoint $T_2^{\dag}$, respectively. One of the axioms characterizing this construction was referred to as `associativity', as such a property renders an unambiguous extension of the construction to 3-time quantum dynamics $(\rho,\E,\F)$. The topic of associativity was taken up in detail in Ref.~\cite{Fu_2025}, where it was shown that associativity indeed yields an unambiguous extension of states over time to $N$-time quantum dynamics for all $N>1$. While a formula for the general $N$-time KD state over time was not given in Ref.~\cite{Fu_2025}, the 3-time formula $T_3=(\rho_1 \otimes \one_2 \otimes \one_3)(\J[\E] \otimes \one_3)(\one_1 \otimes \J[\F])$ appears in the proof of Proposition~65, where the KD state over time is referred to as the `right bloom' construction. In Ref.~\cite{Lie_2025}, the KD state over time was referred to as the `left product state over time', and a recursive formula for the general $N$-time case was given in terms of $T_2$. The KD state over time also plays a prominent role in Ref.~\cite{Lie_2026}, where an operational interpretation of KD states over time (also referred to as the `left product' therein) was established in terms of the interference terms in the probabilities associated with interferometric probes of $N$-time quantum dynamics.
\er

We now state and prove several basic properties of KD states over time. While some of these properties are well known, we include all proofs here to keep the presentation self-contained.

\bn[Basic properties of KD states over time]\label{prop:kd-properties}
Let $(\rho_1,\E_1,\ldots,\E_{N-1})$ be an $N$-time quantum dynamics with associated KD state over time $T_N = T_N(\rho_1,\E_1,\ldots,\E_{N-1}) \in \Lin(\H_{1\cdots N})$. For each $m \in \{1,\ldots,N\}$, let $\mathrm{Tr}_m$ denote the partial trace over $\H_m$, and let $\mathrm{Tr}_{\widehat{m}}$ denote the partial trace over all tensor factors other than $\H_m$. Then $T_N$ satisfies the following properties.
\begin{enumerate}[(i)]
\item \label{item:kd-correlators} \textbf{Kirkwood-Dirac correlators.} For all operators $\O_k \in \Lin(\H_k)$ ($k=1,\ldots,N$),
\be\label{eq:kd-correlator}
\mathrm{Tr}\big[T_N(\O_1 \otimes \cdots \otimes \O_N)\big] = \mathrm{Tr}\Big(\O_N \E_{N-1}\big(\cdots \O_2 \E_1(\O_1 \rho_1)\cdots\big)\Big)\, .
\ee
In particular, when each $\O_k$ is an orthogonal projector (or POVM element), this trace pairing yields the $N$-time Kirkwood-Dirac quasiprobability distribution.

\item \label{item:kd-marginals} \textbf{Unit trace and marginal consistency.} One has $\mathrm{Tr}(T_N) = 1$. For $N = 2$, the partial traces satisfy $\mathrm{Tr}_1(T_2) = \E_1(\rho_1)$ and $\mathrm{Tr}_2(T_2) = \rho_1$. For $N > 2$, tracing out any single factor $\H_m$ ($m \in \{1,\ldots,N\}$) yields the reduced $(N-1)$-time KD state over time
\be\label{eq:kd-kolmogorov}
\mathrm{Tr}_m(T_N) =
\begin{cases}
T_{N-1}(\E_1(\rho_1),\E_2,\ldots,\E_{N-1}) & \text{if } m = 1\, , \\
T_{N-1}(\rho_1,\E_1,\ldots,\E_m \circ \E_{m-1},\ldots,\E_{N-1}) & \text{if } 1 < m < N\, , \\
T_{N-1}(\rho_1,\E_1,\ldots,\E_{N-2}) & \text{if } m = N\, .
\end{cases}
\ee
Consequently, for every $m \in \{1,\ldots,N\}$, the single-time marginal on $\H_m$ recovers the evolved density operator $\rho_m = (\E_{m-1} \circ \cdots \circ \E_1)(\rho_1)$ (with $\rho_1$ unchanged for $m=1$), i.e.,
\be\label{eq:kd-single-marginal}
\mathrm{Tr}_{\widehat{m}}(T_N) = \rho_m\, .
\ee

\item \label{item:kd-convex} \textbf{Convex linearity and multilinearity.} For any finite probability distribution $\{p_\alpha\}$, any density operators $\rho_1^{(\alpha)} \in \Lin(\H_1)$, and any quantum channels $\E_m^{(\alpha)} \colon \Lin(\H_m) \to \Lin(\H_{m+1})$ at a fixed step $m \in \{1,\ldots,N-1\}$,
\be\label{eq:kd-convex-state}
T_N\!\left(\sum_\alpha p_\alpha \rho_1^{(\alpha)},\E_1,\ldots,\E_{N-1}\right) = \sum_\alpha p_\alpha T_N\!\left(\rho_1^{(\alpha)},\E_1,\ldots,\E_{N-1}\right)
\ee
and
\be\label{eq:kd-convex-channel}
T_N\!\left(\rho_1,\E_1,\ldots,\sum_\alpha p_\alpha \E_m^{(\alpha)},\ldots,\E_{N-1}\right) = \sum_\alpha p_\alpha T_N\!\left(\rho_1,\E_1,\ldots,\E_m^{(\alpha)},\ldots,\E_{N-1}\right)\, .
\ee

\item \label{item:kd-unitary-cov} \textbf{Local unitary covariance.} For any unitary operators $V_k \in \Lin(\H_k)$ ($k=1,\ldots,N$), writing $\Ad_{V_k}(X) = V_k X V_k^\dagger$, one has
\be\label{eq:kd-unitary-cov}
T_N\big(\Ad_{V_1}(\rho_1),\Ad_{V_2}\circ\E_1\circ\Ad_{V_1^\dagger},\ldots,\Ad_{V_N}\circ\E_{N-1}\circ\Ad_{V_{N-1}^\dagger}\big) = \left(\bigotimes_{k=1}^{N} V_k\right) T_N \left(\bigotimes_{k=1}^{N} V_k^\dagger\right)\, .
\ee

\item \label{item:kd-parallel} \textbf{Parallel composition.} Given a second $N$-time quantum dynamics $(\sigma_1,\F_1,\ldots,\F_{N-1})$ on finite-dimensional Hilbert spaces $\H'_1,\ldots,\H'_N$, let $\Phi \colon \bigotimes_{k=1}^{N}(\H_k \otimes \H'_k) \to \H_{1\cdots N} \otimes \H'_{1\cdots N}$ denote the canonical unitary reordering of tensor factors. Then
\be\label{eq:kd-parallel}
\Phi\, T_N(\rho_1 \otimes \sigma_1,\E_1 \otimes \F_1,\ldots,\E_{N-1} \otimes \F_{N-1})\,\Phi^\dagger = T_N(\rho_1,\E_1,\ldots,\E_{N-1}) \otimes T_N(\sigma_1,\F_1,\ldots,\F_{N-1})\, .
\ee

\item \label{item:kd-norm} \textbf{Operator-norm and spectral contraction.} With $\|\cdot\|_\infty$ denoting the operator norm,
\be\label{eq:kd-norm}
\|T_N\|_\infty \le \|\rho_1\|_\infty \le 1\, .
\ee
In particular, every singular value of $T_N$ lies in $[0,1]$, and every eigenvalue $\lambda$ of $T_N$ lies in the closed unit disk.
\end{enumerate}
\en

\bprf
\noindent\textbf{Item~\ref{item:kd-correlators}.} We proceed by induction on $N > 1$. For the base case $N = 2$, let $\{\ket{i}\}$ be an orthonormal basis of $\H_1$. Expanding $T_2 = (\rho_1 \otimes \one_2)\J[\E_1] = \sum_{i,j} \rho_1 \dyad{i}{j} \otimes \E_1(\dyad{j}{i})$ yields
\begin{align*}
\mathrm{Tr}\big[T_2(\O_1 \otimes \O_2)\big]
&= \sum_{i,j} \mathrm{Tr}(\rho_1 \dyad{i}{j}\O_1)\,\mathrm{Tr}\big(\E_1(\dyad{j}{i})\O_2\big) \\
&= \sum_{i,j} \bra{j}\O_1 \rho_1\ket{i}\,\mathrm{Tr}\big(\O_2\E_1(\dyad{j}{i})\big) \\
&= \mathrm{Tr}\Big(\O_2 \E_1\Big(\sum_{i,j} \bra{j}\O_1 \rho_1\ket{i}\dyad{j}{i}\Big)\Big) \\
&= \mathrm{Tr}\big(\O_2 \E_1(\O_1 \rho_1)\big)\, ,
\end{align*}
where the last equality uses the completeness identity $\sum_{i,j} \bra{j}X\ket{i}\dyad{j}{i} = X$ with $X = \O_1 \rho_1$. Now assume that~\eqref{eq:kd-correlator} holds for a given integer $N > 1$, and consider an $(N+1)$-time quantum dynamics $(\rho_1,\E_1,\ldots,\E_N)$. By Definition~\ref{def:kd}, we may write $T_{N+1} = (T_N \otimes \one_{N+1})\J_{N,N+1}[\E_N]$, where $T_N = T_N(\rho_1,\E_1,\ldots,\E_{N-1}) \in \Lin(\H_{1\cdots N})$. Letting $\{\ket{i}\}$ be an orthonormal basis of $\H_N$ so that $\J_{N,N+1}[\E_N] = \sum_{i,j} \one_{1\cdots(N-1)} \otimes \dyad{i}{j} \otimes \E_N(\dyad{j}{i})$, we obtain
\begin{align*}
\mathrm{Tr}\big[T_{N+1}(\O_1 \otimes \cdots \otimes \O_{N+1})\big]
&= \sum_{i,j} \mathrm{Tr}\Big[T_N\big(\O_1 \otimes \cdots \otimes \O_{N-1} \otimes \dyad{i}{j}\O_N\big)\Big]\,\mathrm{Tr}\big(\O_{N+1}\E_N(\dyad{j}{i})\big) \\
&= \sum_{i,j} \mathrm{Tr}\Big(\dyad{i}{j}\O_N \E_{N-1}\big(\cdots \O_2 \E_1(\O_1 \rho_1)\cdots\big)\Big)\,\mathrm{Tr}\big(\O_{N+1}\E_N(\dyad{j}{i})\big) \\
&= \sum_{i,j} \bra{j}\O_N \E_{N-1}\big(\cdots \O_2 \E_1(\O_1 \rho_1)\cdots\big)\ket{i}\,\mathrm{Tr}\big(\O_{N+1}\E_N(\dyad{j}{i})\big) \\
&= \mathrm{Tr}\Big(\O_{N+1}\E_N\big(\O_N \E_{N-1}(\cdots \O_2 \E_1(\O_1 \rho_1)\cdots)\big)\Big)\, ,
\end{align*}
where the second equality applies the induction hypothesis with $\O_N$ replaced by $\dyad{i}{j}\O_N$, and the final equality uses the linearity of $\E_N$ and the trace together with completeness on $\H_N$. This completes the induction and establishes~\eqref{eq:kd-correlator}.

\medskip
\noindent\textbf{Item~\ref{item:kd-marginals}.} Setting $\O_k = \one_k$ for all $k \in \{1,\ldots,N\}$ in~\eqref{eq:kd-correlator} and repeatedly using the trace-preserving property of each channel $\E_k$ gives $\mathrm{Tr}(T_N) = \mathrm{Tr}\big(\E_{N-1}(\cdots \E_1(\rho_1)\cdots)\big) = \mathrm{Tr}(\rho_1) = 1$. Next, fix $m \in \{1,\ldots,N\}$ and arbitrary operators $\O_k \in \Lin(\H_k)$ for $k \ne m$, and write $\O_{\widehat{m}} = \bigotimes_{k \ne m} \O_k$. Setting $\O_m = \one_m$ in~\eqref{eq:kd-correlator} gives
\[
\mathrm{Tr}\big[\mathrm{Tr}_m(T_N)\,\O_{\widehat{m}}\big] = \mathrm{Tr}\big[T_N(\O_1 \otimes \cdots \otimes \O_{m-1} \otimes \one_m \otimes \O_{m+1} \otimes \cdots \otimes \O_N)\big]\, .
\]
For $m = 1$, the right-hand side evaluates to $\mathrm{Tr}\big(\O_N \E_{N-1}(\cdots \O_2 \E_1(\rho_1)\cdots)\big)$, which equals $\mathrm{Tr}(\O_2 \E_1(\rho_1))$ when $N=2$ and equals $\mathrm{Tr}\big[T_{N-1}(\E_1(\rho_1),\E_2,\ldots,\E_{N-1})\,\O_{\widehat{1}}\big]$ when $N > 2$ by Item~\ref{item:kd-correlators}. For $1 < m < N$, the identity $\E_m(\one_m \E_{m-1}(X)) = (\E_m \circ \E_{m-1})(X)$ identifies the right-hand side with $\mathrm{Tr}\big[T_{N-1}(\rho_1,\E_1,\ldots,\E_m \circ \E_{m-1},\ldots,\E_{N-1})\,\O_{\widehat{m}}\big]$. For $m = N$, the trace-preserving property of $\E_{N-1}$ reduces the right-hand side to $\mathrm{Tr}\big(\O_{N-1}\E_{N-2}(\cdots \O_2\E_1(\O_1 \rho_1)\cdots)\big)$, which equals $\mathrm{Tr}(\O_1 \rho_1)$ when $N=2$ and equals $\mathrm{Tr}\big[T_{N-1}(\rho_1,\E_1,\ldots,\E_{N-2})\,\O_{\widehat{N}}\big]$ when $N > 2$. Since product operators $\O_{\widehat{m}}$ span $\Lin\big(\bigotimes_{k \ne m}\H_k\big)$ and the trace pairing is nondegenerate, both the $N=2$ marginal relations and~\eqref{eq:kd-kolmogorov} follow. Finally, setting $\O_k = \one_k$ for all $k \ne m$ in~\eqref{eq:kd-correlator} and using the trace-preserving property of $\E_m,\ldots,\E_{N-1}$ gives
\[
\mathrm{Tr}\big[\mathrm{Tr}_{\widehat{m}}(T_N)\,\O_m\big] = \mathrm{Tr}\Big(\E_{N-1}\big(\cdots \E_m(\O_m \rho_m)\cdots\big)\Big) = \mathrm{Tr}(\O_m \rho_m)
\]
for all $\O_m \in \Lin(\H_m)$, establishing~\eqref{eq:kd-single-marginal}.

\medskip
\noindent\textbf{Item~\ref{item:kd-convex}.} By Definition~\ref{def:lpsot}, the Jamio{\l}kowski assignment $\E \mapsto \J[\E]$ is linear, so $\J_{m,m+1}\big[\sum_\alpha p_\alpha \E_m^{(\alpha)}\big] = \sum_\alpha p_\alpha \J_{m,m+1}[\E_m^{(\alpha)}]$. Substituting $\sum_\alpha p_\alpha \rho_1^{(\alpha)}$ or $\sum_\alpha p_\alpha \E_m^{(\alpha)}$ into Definition~\ref{def:kd} and expanding by the linearity of tensor products and operator multiplication immediately yields~\eqref{eq:kd-convex-state} and~\eqref{eq:kd-convex-channel}.

\medskip
\noindent\textbf{Item~\ref{item:kd-unitary-cov}.} Let $\tilde{T}_N$ denote the left-hand side of~\eqref{eq:kd-unitary-cov}, and let $\O_k \in \Lin(\H_k)$ ($k=1,\ldots,N$) be arbitrary operators. Applying Item~\ref{item:kd-correlators} to $\tilde{T}_N$, cancelling adjacent factors $V_k^\dagger V_k = \one_k$, and using cyclicity of the trace to bring $V_N^\dagger$ to the left of $\O_N$ gives
\begin{align*}
\mathrm{Tr}\big[\tilde{T}_N(\O_1 \otimes \cdots \otimes \O_N)\big]
&= \mathrm{Tr}\Big(\O_N V_N \E_{N-1}\big(V_{N-1}^\dagger \O_{N-1} V_{N-1} \cdots \E_1(V_1^\dagger \O_1 V_1 \rho_1)\cdots\big) V_N^\dagger\Big) \\
&= \mathrm{Tr}\Big((V_N^\dagger \O_N V_N) \E_{N-1}\big(\cdots (V_2^\dagger \O_2 V_2)\E_1((V_1^\dagger \O_1 V_1)\rho_1)\cdots\big)\Big) \\
&= \mathrm{Tr}\Big[T_N\big((V_1^\dagger \O_1 V_1) \otimes \cdots \otimes (V_N^\dagger \O_N V_N)\big)\Big] \\
&= \mathrm{Tr}\left[\left(\bigotimes_{k=1}^{N} V_k\right) T_N \left(\bigotimes_{k=1}^{N} V_k^\dagger\right)(\O_1 \otimes \cdots \otimes \O_N)\right]\, .
\end{align*}
Since product operators span $\Lin(\H_{1\cdots N})$ and the trace pairing is nondegenerate,~\eqref{eq:kd-unitary-cov} follows.

\medskip
\noindent\textbf{Item~\ref{item:kd-parallel}.} Let $T_N^{\mathrm{par}} = T_N(\rho_1 \otimes \sigma_1,\E_1 \otimes \F_1,\ldots,\E_{N-1} \otimes \F_{N-1})$. For any operators $\O_k \in \Lin(\H_k)$ and $\O'_k \in \Lin(\H'_k)$ ($k=1,\ldots,N$), the unitary reordering $\Phi$ satisfies $\Phi^\dagger \big((\O_1 \otimes \cdots \otimes \O_N) \otimes (\O'_1 \otimes \cdots \otimes \O'_N)\big)\Phi = \bigotimes_{k=1}^{N}(\O_k \otimes \O'_k)$. Using cyclicity of the trace and Item~\ref{item:kd-correlators} then gives
\begin{align*}
&\mathrm{Tr}\Big[\Phi\, T_N^{\mathrm{par}}\,\Phi^\dagger \big((\O_1 \otimes \cdots \otimes \O_N) \otimes (\O'_1 \otimes \cdots \otimes \O'_N)\big)\Big] \\
&\qquad = \mathrm{Tr}\left[T_N^{\mathrm{par}} \bigotimes_{k=1}^{N} (\O_k \otimes \O'_k)\right] \\
&\qquad = \mathrm{Tr}\Big((\O_N \otimes \O'_N)(\E_{N-1} \otimes \F_{N-1})\big(\cdots (\O_1 \otimes \O'_1)(\rho_1 \otimes \sigma_1)\cdots\big)\Big) \\
&\qquad = \mathrm{Tr}\Big(\O_N \E_{N-1}(\cdots \O_1 \rho_1 \cdots)\Big)\,\mathrm{Tr}\Big(\O'_N \F_{N-1}(\cdots \O'_1 \sigma_1 \cdots)\Big) \\
&\qquad = \mathrm{Tr}\Big[\big(T_N(\rho_1,\E_1,\ldots,\E_{N-1}) \otimes T_N(\sigma_1,\F_1,\ldots,\F_{N-1})\big)\big((\O_1 \otimes \cdots \otimes \O_N) \otimes (\O'_1 \otimes \cdots \otimes \O'_N)\big)\Big]\, .
\end{align*}
Since such product operators span $\Lin(\H_{1\cdots N} \otimes \H'_{1\cdots N})$ and the trace pairing is nondegenerate,~\eqref{eq:kd-parallel} holds.

\medskip
\noindent\textbf{Item~\ref{item:kd-norm}.} We first show that $\|\J[\E]\|_\infty \le 1$ for any quantum channel $\E \colon \Lin(\H_k) \to \Lin(\H_{k+1})$. By Stinespring's dilation theorem, there exist a finite-dimensional auxiliary Hilbert space $\H_E$ and an isometry $V \colon \H_k \to \H_{k+1} \otimes \H_E$ ($V^\dagger V = \one_k$) such that $\E(X) = \mathrm{Tr}_E(V X V^\dagger)$ for all $X \in \Lin(\H_k)$. Choosing orthonormal bases $\{\ket{a}\}$ of $\H_{k+1}$ and $\{\ket{e}\}$ of $\H_E$, define $W \in \Lin(\H_{k+1} \otimes \H_E \otimes \H_{k+1})$ by
\[
W = \sum_{a,b,e} \dyad{a}{b} \otimes \dyad{e}{e} \otimes \dyad{b}{a}\, .
\]
On product basis vectors, $W(\ket{a} \otimes \ket{e} \otimes \ket{b}) = \ket{b} \otimes \ket{e} \otimes \ket{a}$, so $W^\dagger = W$ and $W^2 = \one$, whence $W$ is unitary and $\|W\|_\infty = 1$. Expanding in an orthonormal basis $\{\ket{i}\}$ of $\H_k$ yields
\begin{align*}
(V^\dagger \otimes \one_{k+1}) W (V \otimes \one_{k+1})
&= \sum_{i,j} \sum_{a,b,e} \dyad{i}{i} V^\dagger (\dyad{a}{b} \otimes \dyad{e}{e}) V \dyad{j}{j} \otimes \dyad{b}{a} \\
&= \sum_{i,j} \dyad{i}{j} \otimes \sum_{a,b,e} \bra{b}\bra{e}V\ket{j}\bra{i}V^\dagger\ket{a}\ket{e}\dyad{b}{a} \\
&= \sum_{i,j} \dyad{i}{j} \otimes \sum_{e} (\one_{k+1} \otimes \bra{e}) V \dyad{j}{i} V^\dagger (\one_{k+1} \otimes \ket{e}) \\
&= \sum_{i,j} \dyad{i}{j} \otimes \mathrm{Tr}_E(V\dyad{j}{i}V^\dagger) \\
&= \J[\E]\, .
\end{align*}
Since $V$ is an isometry, $\|V\|_\infty = \|V^\dagger\|_\infty = 1$, and submultiplicativity of the operator norm gives
\[
\|\J[\E]\|_\infty = \|(V^\dagger \otimes \one_{k+1}) W (V \otimes \one_{k+1})\|_\infty \le \|V^\dagger \otimes \one_{k+1}\|_\infty \|W\|_\infty \|V \otimes \one_{k+1}\|_\infty = 1\, .
\]
Hence $\|\J_{k,k+1}[\E_k]\|_\infty = \|\J[\E_k]\|_\infty \le 1$ for every $k \in \{1,\ldots,N-1\}$. Applying submultiplicativity of the operator norm to Definition~\ref{def:kd} together with $\|\rho_1 \otimes \one_2 \otimes \cdots \otimes \one_N\|_\infty = \|\rho_1\|_\infty \le 1$ yields
\[
\|T_N\|_\infty \le \|\rho_1 \otimes \one_2 \otimes \cdots \otimes \one_N\|_\infty \prod_{k=1}^{N-1} \|\J_{k,k+1}[\E_k]\|_\infty \le \|\rho_1\|_\infty \le 1\, ,
\]
establishing~\eqref{eq:kd-norm}. Consequently, every singular value of $T_N$ is bounded above by $\|T_N\|_\infty \le 1$, and for any eigenvalue $\lambda$ of $T_N$ with unit eigenvector $\ket{v}$, one has $|\lambda| = \|T_N\ket{v}\| \le \|T_N\|_\infty \le 1$.
\eprf

Taken together, the properties in Proposition~\ref{prop:kd-properties} show that $T_N$ shares much of the operational structure of an ordinary multipartite density operator while isolating where inherently temporal features appear. In particular, Item~\ref{item:kd-correlators} plays the role of a multi-time Born rule, with the trace pairing~\eqref{eq:kd-correlator} generating $N$-time Kirkwood-Dirac (KD) quasiprobability distributions when evaluated on projective measurements or POVMs. Although introduced nearly a century ago~\cite{Kirkwood_1933,Dirac_1945}, there has been a resurgence of interest in the KD distribution across various subfields within quantum physics~\cite{Lostaglio_2023,Arvidsson_Shukur_2024}. In the study of weak values and weak-value amplification, KD quasiprobabilities provide the natural statistical weights underlying post-selected expectation values, direct quantum state measurements, and interferometric probing schemes~\cite{Lostaglio_2023,Arvidsson_Shukur_2024}. In quantum thermodynamics, they offer a principled framework for characterizing the statistics of incompatible observables, such as work, heat, and energy fluctuations, in regimes where conventional two-point projective measurements erase initial quantum coherence~\cite{Lostaglio_2023,Arvidsson_Shukur_2024}. More directly, the nonreality and negativity of the KD distribution have also been used to construct faithful, operationally accessible quantifiers of quantum coherence relative to a reference basis, connecting KD nonclassicality to measurement uncertainty and quantum contextuality~\cite{Budiyono_2023,Budiyono_2024}. Across these settings, negative or non-real values of the KD distribution serve as direct signatures of nonclassicality, enabling metrological and thermodynamic advantages that cannot be reproduced by classical stochastic models. Encoding all multi-time KD correlators within a single operator $T_N$ thus elevates the KD quasiprobability formalism to an operator-level description of quantum dynamics.

The remaining properties in Proposition~\ref{prop:kd-properties} clarify how $T_N$ behaves under standard operational manipulations. Item~\ref{item:kd-marginals} functions as a quantum analog of Kolmogorov consistency for classical stochastic processes. Tracing out an intermediate time slice simply composes the intervening channels, while tracing out the initial or final time advances the state or discards the subsequent evolution in accordance with causality. Repeated marginalization to any single slice then recovers the usual evolved density operator $\rho_m$. Item~\ref{item:kd-convex} reflects compatibility with classical statistical mixtures, ensuring that ignorance about either the initial preparation or the channel implemented at any step corresponds directly to the associated convex combination of KD states over time. Items~\ref{item:kd-unitary-cov} and~\ref{item:kd-parallel} encode basic structural symmetries, confirming that local changes of basis at each time slice act covariantly on $T_N$ and that independent processes evolving in parallel combine via the tensor product. Finally, Item~\ref{item:kd-norm} establishes a natural spectral stability bound. Although non-Hermiticity allows $T_N$ to acquire complex eigenvalues and precludes positive semi-definiteness, its operator norm cannot exceed $\|\rho_1\|_\infty \le 1$, confining all singular values to $[0,1]$ and all eigenvalues to the closed unit disk. This bound generalizes the $[0,1]$ spectral range of ordinary density operators and provides a natural starting point for the study of information measures for non-Hermitian spacetime states.

\br[Pseudo-density matrices] \label{PDMs}
Given a $2$-time quantum dynamics $(\rho_1,\E)$, the Hermitian part $H_2=\frac{1}{2}(T_2+T_2^{\dag})$ of the KD state over time $T_2$ is the \emph{pseudo-density matrix} (PDM) associated with $(\rho_1,\E)$~\cite{FJV15,FuPa24}. Operationally, a PDM associated with $(\rho_1,\E)$ encodes correlations between sequential measurements of a quantum system prepared in state $\rho_1$ with the quantum channel $\E$ modeling the evolution of the system between measurements. In particular, if $\{\Pi_i\}$ and $\{\Xi_j\}$ are projective measurements on the input and output of $\E$, respectively, then the real-valued quasi-probability distribution $Q(i,j)=\Tr[H_2(\Pi_i\otimes \Xi_j)]$ is the temporal Margenau-Hill distribution associated with a sequential measurement of $\{\Pi_i\}$ followed by $\{\Xi_j\}$~\cite{FuWu_2026}. Following Ref.~\cite{Johansen_2007}, the distribution $Q(i,j)$ may be viewed as the joint probability distribution associated with such a sequential measurement modified by a term quantifying the disturbance of $\rho_1$ due to the initial measurement $\{\Pi_i\}$. If the projective measurements $\{\Pi_i\}$ and $\{\Xi_j\}$ are in fact binary projective measurements, and we form the dichotomic observables $\O_1=\Pi_1-\Pi_2$ and $\O_2=\Xi_1-\Xi_2$, then the two-time expectation value $\langle \O_1,\O_2\rangle$ corresponding to the product of a sequential measurement of $\O_1$ followed by $\O_2$ is given by 
\[
\langle \O_1,\O_2\rangle=\Tr(\E(\Pi_1\rho_1\Pi_1)\O_2)-\Tr(\E(\Pi_2\rho_1\Pi_2)\O_2)\, .
\]
It is then straightforward to show that the PDM $H_2$ satisfies $\langle \O_1,\O_2\rangle=\Tr(H_2\O_1\otimes \O_2)$, which is analogous to the way in which KD states over time encode Kirkwood-Dirac correlators (cf. \eqref{eq:kd-correlator}). While PDMs naturally extend to $N$-time dynamics $(\rho_1,\E_1,\ldots,\E_{N-1})$, for $N>2$ the associated PDM $H_N$ is no longer the Hermitian part of the KD state over time $T_N$. Specifically, it was shown in Refs.~\cite{Fu_2025,Liu_2025} that the $N$-time PDM $H_N$ associated with an $N$-time dynamics $(\rho_1,\E_1,\ldots,\E_{N-1})$ may be given by the formula 
\[
H_N=\mathbf{Jor}\big(\rho_1\otimes \mathds{1}_2\otimes \cdots \otimes \mathds{1}_N,\J_{12}[\E_1],\ldots,\J_{N-1,N}[\E_{N-1}]\big)\, ,
\]
where $\mathbf{Jor}(\cdot,\ldots,\cdot)$ denotes the iterated Jordan product (e.g., $\mathbf{Jor}(X,Y)=\frac{1}{2}(XY+YX)$ and $\mathbf{Jor}(X,Y,Z)=\mathbf{Jor}(\mathbf{Jor}(X,Y),Z)$). PDMs have found numerous applications in quantum information, including the study of quantum channel capacity~\cite{Pisar19}, temporal teleportation~\cite{MVVAPGDG21}, black hole information~\cite{Marletto_2020}, quantum causal inference~\cite{Liu_2025,Liu_2024,Song_2025,Tolotti_2026,Fullwood_Ye_2026}, time-reversal symmetry~\cite{FuPa_2024}, optimal quantum transport~\cite{Riera_2025}, information measures~\cite{FuPa_2023,FuWu_2024}, testing macroscopic realism~\cite{Comar_2026,Bartkiewicz_2026}, and quantum thermodynamics~\cite{Stamatova_2025}.
\er

\section{Spacetime states}\label{sec:equiv}

In standard quantum mechanics, space and time play fundamentally asymmetric roles, with spatial subsystems represented by tensor factors of a Hilbert space while time enters only as an external classical parameter. To formulate quantum theory in a spacetime-symmetric operator language that bridges canonical quantization, relativistic field theory, and the path-integral formalism, Diaz and collaborators introduced \emph{spacetime quantum mechanics}~\cite{Diaz_2024,Diaz_2026a,Diaz_2026}. In this framework, an independent copy of the system's Hilbert space is assigned to each time slice, and unitary dynamics is encoded directly on the resulting multi-time Hilbert space through a quantum action operator rather than an external clock. Pairing this quantum action with an initial density operator yields a single spacetime state whose spacetime traces reproduce multi-time correlation functions in direct analogy with ordinary density operators. 

We now define the operators entering the spacetime state formalism for time-homogeneous unitary evolution as developed in Refs.~\cite{Diaz_2026a,Diaz_2026}. In this setting, we identify every factor $\H_k$ with a single fixed finite-dimensional Hilbert space $\H$, so that $\H_{1\cdots N} = \H^{\otimes N}$, and we write $\one \in \Lin(\H)$ for the identity operator on $\H$. Throughout, we set $\hbar = 1$.

\bd[Swap and cyclic shift operators]\label{def:swap}
Given Hilbert spaces $\H$ and $\H'$, the \define{swap operator} is the linear map ${\tt SWAP}:\H\otimes \H'\to \H'\otimes \H$ defined on product vectors by
\[
{\tt SWAP}(\ket{\psi}_{\H} \otimes \ket{\phi}_{\H'}) = \ket{\phi}_{\H'} \otimes \ket{\psi}_{\H}\, .
\]
For each $k \in \{1,\dots,N-1\}$, let ${\tt SWAP}_{k,k+1} \in \Lin(\H^{\otimes N})$ denote the operator acting as ${\tt SWAP}$ on $\H_{k} \otimes \H_{k+1}$ and as the identity on all remaining tensor factors. The \define{cyclic shift operator} is then the operator $C_N \in \Lin(\H^{\otimes N})$ given by
\[
C_N = {\tt SWAP}_{12}{\tt SWAP}_{23}\cdots{\tt SWAP}_{N-1,N}\, .
\]
In particular, $C_2 = {\tt SWAP}$ and
\[
C_3 = {\tt SWAP}_{12}{\tt SWAP}_{23} = ({\tt SWAP}\otimes \one_3)(\one_1\otimes {\tt SWAP})\, .
\]
On product vectors in $\H^{\otimes N}$, the cyclic shift $C_N$ acts by
\be\label{eq:CN-action}
C_N(\ket{\psi_1} \otimes \ket{\psi_2} \otimes \cdots \otimes \ket{\psi_N}) = \ket{\psi_N} \otimes \ket{\psi_1} \otimes \cdots \otimes \ket{\psi_{N-1}}\, .
\ee
\ed

\bd[Spacetime states~\cite{Diaz_2026}]\label{def:spacetime}
Let $H \in \Lin(\H)$ be a Hermitian Hamiltonian, let $\epsilon > 0$ be a time step, and set $U = e^{-i\epsilon H}$. For each integer $N > 1$, let $\mathcal{P}_N \in \Lin(\H^{\otimes N})$ be a Hermitian operator satisfying $e^{i\epsilon \mathcal{P}_N} = C_N$, and define $\mathcal{K}_N \in \Lin(\H^{\otimes N})$ to be the operator given by
\be
\mathcal{K}_N = \sum_{k=1}^{N} \one^{\otimes (k-1)} \otimes H \otimes \one^{\otimes(N-k)}\, ,
\ee
so that $e^{-i\epsilon \mathcal{K}_N} = U^{\otimes N}$. The \define{$N$-time quantum action} is then the operator $\mathcal{S}_N\in \Lin(\H^{\otimes N})$ satisfying
\be \label{QNTUMAXT87}
e^{i\mathcal{S}_N} = \big(e^{iN\epsilon H} \otimes \one^{\otimes(N-1)}\big) C_N e^{-i\epsilon \mathcal{K}_N}\, ,
\ee
and the \define{$N$-time spacetime state} associated with an initial density operator $\rho_1$ is the operator $R_N \in \Lin(\H^{\otimes N})$ given by
\be
R_N = \big(\rho_1 \otimes \one^{\otimes(N-1)}\big) e^{i\mathcal{S}_N}\, .
\ee
\ed

\bx
For $N=2$ and $N=3$ one has $\mathcal{K}_2 = H \otimes \one + \one \otimes H$, $\mathcal{K}_3 = H \otimes \one \otimes \one + \one \otimes H \otimes \one + \one \otimes \one \otimes H$, and
\begin{align*}
e^{i\mathcal{S}_2} &= (e^{2i\epsilon H} \otimes \one_2) C_2 e^{-i\epsilon \mathcal{K}_2}\, , & R_2 &= (\rho_1 \otimes \one_2) e^{i\mathcal{S}_2}\, ,  \\
e^{i\mathcal{S}_3} &= (e^{3i\epsilon H} \otimes \one_2 \otimes \one_3) C_3 e^{-i\epsilon \mathcal{K}_3}\, , & R_3 &= (\rho_1 \otimes \one_2 \otimes \one_3) e^{i\mathcal{S}_3}\, . 
\end{align*}
\ex

\br[Applications and key results of the spacetime state formalism]\label{rem:spacetime-literature}
In Ref.~\cite{Diaz_2026a}, Diaz and Rossignoli develop the program of spacetime states for finite-dimensional systems as well as bosonic and fermionic fields, showing that spacetime traces weighted by the exponential of a quantum action recover conventional unitary dynamics, thermal correlators, and relativistic Feynman propagators, while reproducing fermionic path integrals without Grassmann variables. They introduce non-Hermitian spacetime states through generalized purifications, giving multi-time correlators an interpretation as weak values, and establish a quantum principle of stationary action in which Euclidean spacetime states extremize a pseudo-entropy-based free-energy functional. They also show that second quantization of the Page-Wootters formalism yields spacetime algebras and free relativistic quantum actions, while identifying the two-time operator $T_{AB}$ of~\cite{Milekhin_2025} with the Hermitian adjoint of the bosonic spacetime state.

Building on this foundation, Diaz, Cerezo, and Braccia~\cite{Diaz_2026} place the spacetime state $R_N$ at the center of the theory and use it to relate a wide range of spacetime approaches to quantum mechanics. They show that spacetime states for closed systems and their reductions to spacetime subregions retain key features of ordinary density operators, including vanishing pseudo-entropies for pure closed histories and matching spectra across complementary bipartitions, while their non-Hermiticity captures time ordering and causal influence. Working with spacetime states and their multi-copy extensions (which allow for both forward and backward time evolution as well as out-of-time-order correlators), they demonstrate that path integrals, 2-time quantum states over time~\cite{HHPBS17}, 2-time pseudo-density matrices~\cite{FJV15}, superdensity operators~\cite{Cotler_2017}, Page-Wootters states~\cite{Page_Wootters_1983}, and the 2-time spacetime density matrices of~\cite{Milekhin_2025} all emerge through basis evaluations, partial traces, or linear maps applied to spacetime states. They further apply this framework to quantum foundations and many-body physics, characterizing Leggett-Garg violations through the negativity of the Hermitian part $\frac{1}{2}(R_N + R_N^\dagger)$, probing causal propagation and operator scrambling in spin chains, and connecting spacetime states to tensor-network methods~\cite{Banuls_2009,Lerose_2021}, covariant quantum field theory~\cite{Diaz_2024,Haag_Kastler_1964}, quantum reference frames~\cite{Aharonov_Kaufherr_1984,Giacomini_2019}, and spacetime classical mechanics~\cite{Dirac_1950}.
\er

We now prove that the spacetime states of Diaz et al. are in fact KD states over time.

\blem\label{lem:unitary-jam}
Let $\H$ be a finite-dimensional Hilbert space, let $U \in \Lin(\H)$ be a unitary operator, and let $\E \colon \Lin(\H) \to \Lin(\H)$ be the associated unitary channel given by $\E(\rho) = U\rho U^\dagger$. Then
\be\label{eq:unitary-jam}
\J[\E] = (U^\dagger \otimes U){\tt SWAP}\, .
\ee
\elem

\bprf
Expanding $\J[\E]$ in an orthonormal basis $\{\ket{i}\}$ of $\H$ yields
\begin{align*}
\J[\E] &= \sum_{i,j} \dyad{i}{j} \otimes U\dyad{j}{i}U^\dagger \\
&= (\one \otimes U)\Big(\sum_{i,j} \dyad{i}{j} \otimes \dyad{j}{i}\Big)(\one \otimes U^\dagger) \\
&= (\one \otimes U)\,{\tt SWAP}\,(\one \otimes U^\dagger) \\
&=(U^\dagger \otimes U){\tt SWAP}\, ,
\end{align*}
as desired.
\eprf

\bt \label{prop:N-slices}
Let $H \in \Lin(\H)$ be a Hermitian Hamiltonian, let $\epsilon > 0$, let $U = e^{-i\epsilon H}$, and let $\E \colon \Lin(\H) \to \Lin(\H)$ be the associated unitary channel given by $\E(\rho) = U\rho U^\dagger$. Then for every integer $N > 1$, we have
\be\label{eq:N-slice-jam}
e^{i\mathcal{S}_N} = \big((U^\dagger)^{N-1} \otimes U^{\otimes(N-1)}\big) C_N = \prod_{k=1}^{N-1} \J_{k,k+1}[\E]\, .
\ee
Consequently, for any initial density operator $\rho_1 \in \Lin(\H)$, the $N$-time spacetime state $R_N$ coincides with the KD state over time $T_N$ associated with the $N$-time quantum dynamics $(\rho_1,\E,\ldots,\E)$, i.e.,
\be\label{eq:RN-TN}
R_N = T_N\, .
\ee
\et

\bprf
We first observe that for any linear operators $A_1, \ldots, A_N \in \Lin(\H)$ and any product vector $\ket{\psi_1} \otimes \cdots \otimes \ket{\psi_N} \in \H^{\otimes N}$, applying~\eqref{eq:CN-action} yields
\begin{align*}
C_N(A_1 \otimes \cdots \otimes A_N)(\ket{\psi_1} \otimes \cdots \otimes \ket{\psi_N})
&= C_N(A_1\ket{\psi_1} \otimes \cdots \otimes A_N\ket{\psi_N}) \\
&= A_N\ket{\psi_N} \otimes A_1\ket{\psi_1} \otimes \cdots \otimes A_{N-1}\ket{\psi_{N-1}} \\
&= (A_N \otimes A_1 \otimes \cdots \otimes A_{N-1}) C_N(\ket{\psi_1} \otimes \cdots \otimes \ket{\psi_N})\, .
\end{align*}
Since product vectors span $\H^{\otimes N}$, we obtain the intertwining identity
\be\label{eq:CN-intertwine}
C_N(A_1 \otimes A_2 \otimes \cdots \otimes A_N) = (A_N \otimes A_1 \otimes \cdots \otimes A_{N-1}) C_N\, .
\ee

To establish the first equality in~\eqref{eq:N-slice-jam}, set $A_1 = \cdots = A_N = U$ in~\eqref{eq:CN-intertwine}, which shows that $C_N$ commutes with $e^{-i\epsilon \mathcal{K}_N} = U^{\otimes N}$. Using Definition~\ref{def:spacetime} together with $e^{iN\epsilon H} = (U^\dagger)^N$ and $U^\dagger U = \one$, we have
\begin{align*}
e^{i\mathcal{S}_N} &= \big(e^{iN\epsilon H} \otimes \one^{\otimes(N-1)}\big) C_N e^{-i\epsilon \mathcal{K}_N} \\
&= \big((U^\dagger)^N \otimes \one^{\otimes(N-1)}\big) C_N U^{\otimes N} \\
&= \big((U^\dagger)^N \otimes \one^{\otimes(N-1)}\big) U^{\otimes N} C_N \\
&= \big((U^\dagger)^N U \otimes U^{\otimes(N-1)}\big) C_N \\
&= \big((U^\dagger)^{N-1} \otimes U^{\otimes(N-1)}\big) C_N\, .
\end{align*}

We next prove the second equality in~\eqref{eq:N-slice-jam} by induction on $N > 1$. For the base case $N = 2$, Definition~\ref{def:swap} gives $C_2 = {\tt SWAP}$, so Lemma~\ref{lem:unitary-jam} directly yields
\[
\J_{12}[\E] = \J[\E] = (U^\dagger \otimes U){\tt SWAP} = (U^\dagger \otimes U) C_2\, ,
\]
thus \eqref{eq:N-slice-jam} holds for $N=2$. Now assume that the equality \eqref{eq:N-slice-jam} holds for a given integer $N > 1$, meaning that
\[
\prod_{k=1}^{N-1} \J_{k,k+1}[\E] = \big((U^\dagger)^{N-1} \otimes U^{\otimes(N-1)}\big) C_N
\]
as operators on $\H^{\otimes N}$. Tensoring on the right with $\one$ embeds this identity into $\Lin(\H^{\otimes(N+1)})$ as
\be \label{TENEQX17}
\prod_{k=1}^{N-1} \J_{k,k+1}[\E] \otimes \mathds{1}= \big((U^\dagger)^{N-1} \otimes U^{\otimes(N-1)} \otimes \one\big)(C_N \otimes \one)\, .
\ee
By Lemma~\ref{lem:unitary-jam}, the operator $\J_{N,N+1}[\E] \in \Lin(\H^{\otimes(N+1)})$ satisfies
\[
\J_{N,N+1}[\E] = \big(\one^{\otimes(N-1)} \otimes U^\dagger \otimes U\big){\tt SWAP}_{N,N+1}\, .
\]
Multiplying both sides of \eqref{TENEQX17} on the right by $\J_{N,N+1}[\E]$ therefore yields
\be\label{eq:ind-step}
\prod_{k=1}^{N} \J_{k,k+1}[\E] = \big((U^\dagger)^{N-1} \otimes U^{\otimes(N-1)} \otimes \one\big)(C_N \otimes \one)\big(\one^{\otimes(N-1)} \otimes U^\dagger \otimes U\big){\tt SWAP}_{N,N+1}\, .
\ee
Moreover, by \eqref{eq:CN-intertwine} we have
\[
(C_N \otimes \one)\big(\one^{\otimes(N-1)} \otimes U^\dagger \otimes U\big) = \big(U^\dagger \otimes \one^{\otimes(N-1)} \otimes U\big)(C_N \otimes \one)\, ,
\]
and by Definition~\ref{def:swap} we have $(C_N \otimes \one){\tt SWAP}_{N,N+1} = {\tt SWAP}_{12}\cdots{\tt SWAP}_{N-1,N}{\tt SWAP}_{N,N+1} = C_{N+1}$. Substituting these two relations into~\eqref{eq:ind-step} then yields
\begin{align*}
\prod_{k=1}^{N} \J_{k,k+1}[\E] &= \big((U^\dagger)^{N-1} \otimes U^{\otimes(N-1)} \otimes \one\big)\big(U^\dagger \otimes \one^{\otimes(N-1)} \otimes U\big)(C_N \otimes \one){\tt SWAP}_{N,N+1} \\
&= \big((U^\dagger)^{N-1} U^\dagger \otimes U^{\otimes(N-1)} \otimes U\big) C_{N+1} \\
&= \big((U^\dagger)^N \otimes U^{\otimes N}\big) C_{N+1}\, ,
\end{align*}
completing the induction and establishing~\eqref{eq:N-slice-jam}.

Finally, given any initial density operator $\rho_1 \in \Lin(\H)$, left-multiplying the outer equality of~\eqref{eq:N-slice-jam} by $\rho_1 \otimes \one^{\otimes(N-1)}$ and invoking Definition~\ref{def:spacetime} together with Definition~\ref{def:kd} (with $\H_k = \H$ and $\E_k = \E$ for all $k$) yields
\[
R_N = \big(\rho_1 \otimes \one^{\otimes(N-1)}\big) e^{i\mathcal{S}_N} = \big(\rho_1 \otimes \one^{\otimes(N-1)}\big) \prod_{k=1}^{N-1} \J_{k,k+1}[\E] = T_N\, ,
\]
which establishes~\eqref{eq:RN-TN}.
\eprf

\br\label{rem:diaz-n2}
For two time slices, Theorem~2 and Section~III.B of Ref.~\cite{Diaz_2026} establish via a Stinespring dilation that the reduced spacetime state associated with an arbitrary quantum channel $\E$ takes the form $(\rho_1 \otimes \one_2)\J[\E]$ (reducing to $R_2 = (\rho_1 \otimes \one_2)\J[\E] = T_2$ in the unitary case), and note this operator as a non-Hermitian choice of quantum state over time. However, both the Jamio{\l}kowski representation and the quantum state over time identification in~\cite{Diaz_2026} are restricted to $N=2$, without connecting this construction to the left-product (KD) states over time studied in the literature or establishing the $N$-time factorization $e^{i\mathcal{S}_N} = \prod_{k=1}^{N-1}\J_{k,k+1}[\E]$ and resulting equivalence $R_N = T_N$ for $N \ge 3$ proved in Theorem~\ref{prop:N-slices}.

Moreover, Ref.~\cite{Diaz_2026} does not establish an explicit $N$-time spacetime state for arbitrary CPTP evolutions beyond $N=2$. In Section~II.B of~\cite{Diaz_2026}, the authors remark that sequential memoryless channels may be treated by introducing a separate ancilla at each time step, without deriving a closed-form expression for $N >2$ in terms of an ordered link-product of Jamio{\l}kowski operators. Even their extension to time-dependent unitary evolution in Appendix~A of~\cite{Diaz_2026} is comparatively involved, relying on time-ordered cumulative evolution operators and conjugated cyclic shifts to define the quantum action. In contrast, Definition~\ref{def:kd} yields a simple closed-form expression for the KD state over time $T_N = (\rho_1 \otimes \one_2 \otimes \cdots \otimes \one_N)\prod_{k=1}^{N-1}\J_{k,k+1}[\E_k]$ for arbitrary sequential CPTP evolutions between time slices, even allowing the Hilbert spaces $\H_k$ to vary from slice to slice. In view of Theorem~\ref{prop:N-slices}, KD states over time immediately generalize the spacetime state construction of Ref.~\cite{Diaz_2026} to arbitrary $N$-time CPTP evolutions, without having to define the auxiliary operators $\mathcal{P}_N$, $\mathcal{K}_N$, and $e^{iN\epsilon H} \otimes \one^{\otimes(N-1)}$ needed for the definition of the quantum action \eqref{QNTUMAXT87}.
\er

\section{Spacetime density matrices} \label{sec:sdm-equiv}

A closely related construction to that of KD states over time, termed the \emph{spacetime density matrix}, was introduced by Milekhin, Adamska, and Preskill~\cite{Milekhin_2025} for two-time quantum dynamics (allowing arbitrary quantum channels, with multi-time extensions illustrated for unitary evolution) and systematically developed for arbitrary $N$-time unitary evolution by Guo~\cite{Guo_2025}. Rather than specifying the operator through a product of Jamio{\l}kowski operators or a quantum action, these works define the spacetime density matrix implicitly as the unique operator on $\H_{1\cdots N}$ whose Hilbert-Schmidt contractions with product observables reproduce $N$-point anti-time-ordered Wightman correlation functions~\cite{Weinberg_1995} (which is similar to the approach of Ref.~\cite{Jia_Xiao_2026}). To distinguish this operator from the KD state over time $T_N$, we denote it by $\Sigma_N$ (corresponding to $T_{AB}$ in~\cite{Milekhin_2025} for the case $N=2$ and $T_{C_0\cdots C_{N-1}}$ in~\cite{Guo_2025} for general $N>1$).

Recall that for any quantum channel $\E \colon \Lin(\H_k) \to \Lin(\H_{k+1})$, its \define{Hilbert-Schmidt adjoint} (or Heisenberg dual) is the unique unital completely positive map $\E^* \colon \Lin(\H_{k+1}) \to \Lin(\H_k)$ satisfying $\mathrm{Tr}\big(\E(X)Y\big) = \mathrm{Tr}\big(X\E^*(Y)\big)$ for all $X \in \Lin(\H_k)$ and $Y \in \Lin(\H_{k+1})$.

\bd[Spacetime density matrices, cf.~\cite{Milekhin_2025,Guo_2025}]\label{def:sdm}
Let $(\rho_1,\E_1,\ldots,\E_{N-1})$ be an $N$-time quantum dynamics. The \define{spacetime density matrix} associated with $(\rho_1,\E_1,\ldots,\E_{N-1})$ is the unique operator $\Sigma_N \in \Lin(\H_{1\cdots N})$ satisfying
\be\label{eq:sdm-def}
\mathrm{Tr}\big[\Sigma_N(\O_1 \otimes \cdots \otimes \O_N)\big] = \mathrm{Tr}\Big(\rho_1 \O_1 \E_1^*\big(\O_2 \E_2^*(\cdots \E_{N-2}^*(\O_{N-1}\E_{N-1}^*(\O_N))\cdots)\big)\Big)
\ee
for all operators $\O_k \in \Lin(\H_k)$ ($k=1,\ldots,N$).
\ed

The uniqueness of $\Sigma_N$ follows from the nondegeneracy of the trace pairing $(X,Y) \mapsto \mathrm{Tr}(XY)$ on $\Lin(\H_{1\cdots N})$ together with the fact that product operators $\O_1 \otimes \cdots \otimes \O_N$ span $\Lin(\H_{1\cdots N})$. Note that Definition~\ref{def:sdm} is more general than the multi-time definition of Guo~\cite{Guo_2025}---which is formulated only for unitary evolution---and may equally be viewed as an $N$-time extension of the two-time channel-based definition of Milekhin, Adamska, and Preskill~\cite{Milekhin_2025}. Indeed, for $N=2$ and a general quantum channel $\E \colon \Lin(\H_1) \to \Lin(\H_2)$, Eq.~\eqref{eq:sdm-def} reduces to
\be
\mathrm{Tr}\big[\Sigma_2(\O_1 \otimes \O_2)\big] = \mathrm{Tr}\big(\rho_1 \O_1 \E^*(\O_2)\big)\, ,
\ee
which recovers the full-system spacetime density matrix of~\cite{Milekhin_2025}. On the other hand, for unitary evolution on a fixed Hilbert space $\H_k = \H$ at times $t_1 < t_2 < \cdots < t_N$, where each channel takes the form $\E_k(\rho) = U_k \rho U_k^\dagger$ with $U_k = U(t_{k+1},t_k)$, each Heisenberg dual $\E_k^*(Y) = U_k^\dagger Y U_k$ is an algebra homomorphism. Setting $V_1 = \one$, $V_k = U_{k-1}\cdots U_1 = U(t_k,t_1)$ for $k \ge 2$, and $\O_k(t_k) = V_k^\dagger \O_k V_k$, the nested composition in~\eqref{eq:sdm-def} simplifies to the $N$-point anti-time-ordered correlator
\be\label{eq:sdm-unitary}
\mathrm{Tr}\big[\Sigma_N(\O_1 \otimes \cdots \otimes \O_N)\big] = \mathrm{Tr}\big(\rho_1 \O_1(t_1)\O_2(t_2)\cdots \O_N(t_N)\big)\, ,
\ee
which is precisely the defining relation of the $N$-slice spacetime density matrix in~\cite{Guo_2025}.

\br[Applications and key results of the spacetime density matrix formalism]\label{rem:sdm-literature}
In Refs.~\cite{Milekhin_2025,Guo_2025}, spacetime density matrices and their partial traces onto time-separated subsystems are developed to probe temporal entanglement, causal influence, and many-body dynamics across quantum information theory and quantum field theory. In Ref.~\cite{Milekhin_2025}, Milekhin, Adamska, and Preskill denote the $2$-time spacetime density matrix by $T_{AB}$ and use it to define measures of entanglement in time through its moments $\mathrm{Tr}(T_{AB}^n)$ and the \emph{entanglement $p$-imagitivity} $\|T_{AB}-T_{AB}^\dagger\|_p$. After establishing basic algebraic and spectral properties of $T_{AB}$, they show that $2$-imagitivity obeys a Lieb-Robinson bound in local lattice models and yields both upper and lower bounds on unequal-time commutators, demonstrating that nonzero $2$-imagitivity is necessary and sufficient for signaling from $A$ to $B$. In relativistic quantum field theory, they show that $\mathrm{Tr}(T_{AB}^n)$ coincides with the analytic continuation of spacelike R\'enyi entropies via time-separated twist-operator correlators, providing a measurable realization of timelike pseudo-entropy. They illustrate this framework across spin chains, free fermions, conformal field theories, and holographic models, while also exploring matrix product operator approximations across four time slices. Finally, they introduce controlled-{\tt SWAP} measurement protocols for $\mathrm{Tr}(T_{AB}^2)$ and $\mathrm{Tr}(T_{AB}T_{AB}^\dagger)$ and implement them on an IBM quantum processor.

In Ref.~\cite{Guo_2025}, Guo extends the formalism systematically to arbitrary $N$-time slices via the operators $T_{C_0\cdots C_{N-1}}$ and their subsystem reductions $T_{A_0\cdots A_{N-1}}$. These operators are represented through Schwinger-Keldysh path integrals with single-branch cuts, and their moments are related to complex-valued R\'enyi and von Neumann entropies via the real-time replica method. Recasting the $2$-time construction within a super-operator framework yields Liouville-von Neumann-type equations of motion and leads naturally to further generalizations. In particular, it is shown that allowing distinct initial and final states connects the formalism to weak values and scattering amplitudes, while applying the super-operator to pairs of density matrices yields characterizations of fidelity and thermalization. A central finding of~\cite{Guo_2025} is that while the moments of the full-system spacetime density matrix are time-independent functions of the initial state and Hilbert-space dimension, the moments of reduced spacetime density matrices depend nontrivially on interactions between subsystems and become time-independent when those couplings vanish. To utilize the reduced spacetime density matrix as a probe of subsystem interactions, short-time and weak-coupling perturbative expansions are developed for its second moments. These methods are then illustrated in a coupled two-qubit model and in thermofield double states dual to eternal black holes and traversable wormholes.
\er

We now prove that the spacetime density matrices of Milekhin, Adamska, Preskill, and Guo are precisely the Hermitian adjoints of KD states over time.

\bt \label{thm:sdm-equiv}
Let $(\rho_1,\E_1,\ldots,\E_{N-1})$ be an $N$-time quantum dynamics, with associated KD state over time $T_N$ and spacetime density matrix $\Sigma_N$. Then
\be\label{eq:sdm-kd-dag}
\Sigma_N = T_N^\dagger = \left(\prod_{k=N-1}^{1} \J_{k,k+1}[\E_k]\right)(\rho_1 \otimes \one_2 \otimes \cdots \otimes \one_N)\, ,
\ee
where the ordered product runs from $k = N-1$ on the left to $k = 1$ on the right. In particular, for $N = 2$ and any quantum channel $\E \colon \Lin(\H_1) \to \Lin(\H_2)$ we have $\Sigma_2 = T_2^\dagger = \J[\E](\rho_1 \otimes \one_2)$, and for time-homogeneous unitary evolution $\E_k(\rho) = U\rho U^\dagger$ with $U = e^{-i\epsilon H}$ we have
\be\label{eq:sdm-RN-dag}
\Sigma_N = T_N^\dagger = R_N^\dagger\, ,
\ee
where $R_N$ is the $N$-time spacetime state of Definition~\ref{def:spacetime}.
\et

\bprf
We first verify the second equality in~\eqref{eq:sdm-kd-dag}. Since each quantum channel $\E_k$ is Hermiticity-preserving, $\E_k(\dyad{j}{i})^\dagger = \E_k(\dyad{i}{j})$ for every pair of basis vectors in $\H_k$, and thus
\be
\J[\E_k]^\dagger = \sum_{i,j} \dyad{j}{i} \otimes \E_k(\dyad{j}{i})^\dagger = \sum_{i,j} \dyad{j}{i} \otimes \E_k(\dyad{i}{j}) = \J[\E_k]\, .
\ee
Hence $\J_{k,k+1}[\E_k]^\dagger = \J_{k,k+1}[\E_k]$ for every $k \in \{1,\ldots,N-1\}$. Combining this with $\rho_1^\dagger = \rho_1$ and taking the Hermitian adjoint of $T_N$ in Definition~\ref{def:kd} yields
\be
T_N^\dagger = \left(\prod_{k=N-1}^{1} \J_{k,k+1}[\E_k]\right)(\rho_1 \otimes \one_2 \otimes \cdots \otimes \one_N)\, .
\ee

We next show by induction on $N > 1$ that $T_N^\dagger$ satisfies~\eqref{eq:sdm-def} for all $\O_k \in \Lin(\H_k)$. For the base case $N = 2$, let $\{\ket{i}\}$ be an orthonormal basis of $\H_1$. Expanding $T_2^\dagger = \J[\E_1](\rho_1 \otimes \one_2) = \sum_{i,j} \dyad{i}{j}\rho_1 \otimes \E_1(\dyad{j}{i})$ gives
\begin{align*}
\mathrm{Tr}\big[T_2^\dagger(\O_1 \otimes \O_2)\big]
&= \sum_{i,j} \mathrm{Tr}\big(\dyad{i}{j}\rho_1 \O_1\big)\,\mathrm{Tr}\big(\E_1(\dyad{j}{i})\O_2\big) \\
&= \sum_{i,j} \bra{j}\rho_1 \O_1\ket{i}\,\mathrm{Tr}\big(\dyad{j}{i}\E_1^*(\O_2)\big) \\
&= \sum_{i,j} \bra{j}\rho_1 \O_1\ket{i}\bra{i}\E_1^*(\O_2)\ket{j} \\
&= \mathrm{Tr}\big(\rho_1 \O_1 \E_1^*(\O_2)\big)\, ,
\end{align*}
establishing the claim for $N = 2$.

Now assume that the identity holds for a given integer $N > 1$, and consider an $(N+1)$-time quantum dynamics $(\rho_1,\E_1,\ldots,\E_N)$. By Definition~\ref{def:kd}, we may write $T_{N+1} = (T_N \otimes \one_{N+1})\J_{N,N+1}[\E_N]$, where $T_N \in \Lin(\H_{1\cdots N})$ is the KD state over time associated with $(\rho_1,\E_1,\ldots,\E_{N-1})$. Taking the Hermitian adjoint and using $\J_{N,N+1}[\E_N]^\dagger = \J_{N,N+1}[\E_N]$ yields
\be
T_{N+1}^\dagger = \J_{N,N+1}[\E_N](T_N^\dagger \otimes \one_{N+1})\, .
\ee
Letting $\{\ket{k}\}$ be an orthonormal basis of $\H_N$ so that $\J_{N,N+1}[\E_N] = \sum_{k,l} \one_{1\cdots(N-1)} \otimes \dyad{k}{l} \otimes \E_N(\dyad{l}{k})$, we compute
\begin{align*}
\mathrm{Tr}\big[T_{N+1}^\dagger(\O_1 \otimes \cdots \otimes \O_{N+1})\big]
&= \sum_{k,l} \mathrm{Tr}\Big[\big(\one_{1\cdots(N-1)} \otimes \dyad{k}{l}\big) T_N^\dagger (\O_1 \otimes \cdots \otimes \O_N)\Big]\,\mathrm{Tr}\big[\E_N(\dyad{l}{k})\O_{N+1}\big] \\
&= \sum_{k,l} \mathrm{Tr}\Big[T_N^\dagger\big(\O_1 \otimes \cdots \otimes \O_{N-1} \otimes \O_N\dyad{k}{l}\big)\Big]\,\bra{k}\E_N^*(\O_{N+1})\ket{l} \\
&= \mathrm{Tr}\Big[T_N^\dagger\big(\O_1 \otimes \cdots \otimes \O_{N-1} \otimes \O_N\E_N^*(\O_{N+1})\big)\Big]\, ,
\end{align*}
where the second equality uses cyclicity of the trace on $\H_{1\cdots N}$ together with $\mathrm{Tr}[\E_N(\dyad{l}{k})\O_{N+1}] = \mathrm{Tr}[\dyad{l}{k}\E_N^*(\O_{N+1})] = \bra{k}\E_N^*(\O_{N+1})\ket{l}$, and the third equality uses linearity of the trace and the expansion $\sum_{k,l} \bra{k}\E_N^*(\O_{N+1})\ket{l}\dyad{k}{l} = \E_N^*(\O_{N+1})$. Applying the induction hypothesis with $\O_N$ replaced by $\O_N\E_N^*(\O_{N+1})$ yields
\be
\mathrm{Tr}\big[T_{N+1}^\dagger(\O_1 \otimes \cdots \otimes \O_{N+1})\big] = \mathrm{Tr}\Big(\rho_1 \O_1 \E_1^*\big(\O_2 \E_2^*(\cdots \E_{N-1}^*(\O_N \E_N^*(\O_{N+1}))\cdots)\big)\Big)\, ,
\ee
which completes the induction. By the uniqueness of $\Sigma_N$ in Definition~\ref{def:sdm}, it follows that $\Sigma_N = T_N^\dagger$.

Finally, for $N = 2$ Eq.~\eqref{eq:sdm-kd-dag} specializes immediately to $\Sigma_2 = T_2^\dagger = \J[\E](\rho_1 \otimes \one_2)$, while for time-homogeneous unitary evolution $\E_k(\rho) = U\rho U^\dagger$ with $U = e^{-i\epsilon H}$, Theorem~\ref{prop:N-slices} gives $R_N = T_N$, and taking Hermitian adjoints yields $\Sigma_N = T_N^\dagger = R_N^\dagger$.
\eprf

\br\label{rem:sdm-corollaries}
Taking the Hermitian adjoint of~\eqref{eq:sdm-kd-dag} shows that the KD state over time $T_N$ coincides with $\Sigma_N^\dagger$, which is precisely the time-ordered spacetime density matrix denoted $T_{AB}^\dagger$ in~\cite{Milekhin_2025} and $T'_{C_0\cdots C_{N-1}} = T_{C_0\cdots C_{N-1}}^\dagger$ in~\cite{Guo_2025}. In the unitary setting of~\eqref{eq:sdm-unitary}, conjugating~\eqref{eq:sdm-unitary} with $\O_k^\dagger$ in place of $\O_k$ yields
\be
\mathrm{Tr}\big[T_N(\O_1 \otimes \cdots \otimes \O_N)\big] = \Big(\mathrm{Tr}\big[\Sigma_N(\O_1^\dagger \otimes \cdots \otimes \O_N^\dagger)\big]\Big)^* = \mathrm{Tr}\big(\rho_1 \O_N(t_N)\cdots \O_2(t_2)\O_1(t_1)\big)\, ,
\ee
confirming that $T_N$ (and $R_N$ in the time-homogeneous case) generates $N$-point time-ordered correlation functions. Moreover, whenever each Hilbert space factorizes as $\H_k = \H_{A_k} \otimes \H_{\overline{A}_k}$, taking the partial trace of $\Sigma_N = T_N^\dagger$ over the complementary factors $\overline{A}_1, \ldots, \overline{A}_N$ directly recovers the reduced spacetime density matrices for time-separated subsystems studied in~\cite{Milekhin_2025,Guo_2025}.

In the unitary setting with two time-separated slices, Lemma~2 and Theorem~11 of Ref.~\cite{Diaz_2026} already establish this relation between the spacetime density matrix of~\cite{Milekhin_2025} and the partial trace of $R_N^\dagger$ over all complementary degrees of freedom, which reduces to $\Sigma_2 = R_2^\dagger$ for the full system at $N=2$. However, although Ref.~\cite{Guo_2025} is cited alongside~\cite{Milekhin_2025} in~\cite{Diaz_2026}, Guo's multi-time spacetime density matrix is not analyzed there, nor is the general $N$-time CPTP equivalence $\Sigma_N = T_N^\dagger$ established. Theorem~\ref{thm:sdm-equiv}, together with Theorem~\ref{prop:N-slices}, extends that two-time unitary identification to arbitrary $N > 1$ and arbitrary sequential quantum channels, placing the constructions of~\cite{Diaz_2026,Milekhin_2025,Guo_2025} within the single framework of KD states over time and their Hermitian adjoints.
\er

\br[Relation to the spatiotemporal Kirkwood-Dirac framework of Ref.~\cite{Jia_Xiao_2026}]\label{rem:jia-xiao}
A broad framework for spatiotemporal quantum states based on Kirkwood-Dirac phase space and Bloch tomography has recently been developed by Jia and Xiao~\cite{Jia_Xiao_2026}, building on Refs.~\cite{JK_2024,JMK_2026}. Their formalism encompasses single- and multi-branch Kirkwood-Dirac temporal states, pseudo-density matrices~\cite{FJV15}, doubled-density operators~\cite{JK_2024}, folded states for out-of-time-ordered correlators, and non-Markovian processes with memory. In Section~4.1 of~\cite{Jia_Xiao_2026}, multi-time states are constructed from Jamio{\l}kowski operators via a \emph{padded temporal link product}, with tensor factors ordered from latest to earliest time. Reversing this temporal ordering to $\H_1 \otimes \cdots \otimes \H_N$, their padded temporal link product reproduces the operator product in Definition~\ref{def:kd}. In particular, the right Kirkwood-Dirac temporal state of~\cite{Jia_Xiao_2026} coincides with $T_N$, while its Hermitian adjoint coincides with $T_N^\dagger$, and hence with the spacetime density matrix $\Sigma_N$ by Theorem~\ref{thm:sdm-equiv}. The KD states over time studied here thus sit within the broader hierarchy of~\cite{Jia_Xiao_2026} as the single-branch time-ordered elements. Consequently, Theorems~\ref{prop:N-slices} and~\ref{thm:sdm-equiv} place both the multi-time spacetime states of~\cite{Diaz_2026} and the multi-time spacetime density matrices of~\cite{Guo_2025} inside that same hierarchy.
\er

\section{An axiomatic derivation of spacetime entropy}\label{sec:entropy}

In his seminal work on information theory~\cite{Shannon_1948}, Shannon derived the classical entropy functional $H(p_1,\ldots,p_n) = -\sum_{i=1}^{n} p_i\log p_i$ from three fundamental assumptions, the third of which has come to be known as \emph{strong additivity}. Explicitly, given a finite probability distribution $(p_1,\ldots,p_k)$ and conditional probability distributions $(q_{i1},\ldots,q_{im_i})$ for $i = 1,\ldots,k$, strong additivity requires that
\be\label{eq:shannon-sa}
H(p_1 q_{11},\ldots,p_1 q_{1m_1},\ldots,p_k q_{k1},\ldots,p_k q_{km_k}) = H(p_1,\ldots,p_k) + \sum_{i=1}^{k} p_i H(q_{i1},\ldots,q_{im_i})\, .
\ee
In the quantum setting, this condition lifts naturally to convex block-diagonal mixtures $\bigoplus_{i=1}^{k} p_i \rho_i$ of density operators, for which the von~Neumann entropy $S(\rho) = -\mathrm{Tr}(\rho\log\rho)$ satisfies $S\big(\bigoplus_{i=1}^{k} p_i \rho_i\big) = S\big(\operatorname{diag}(p_1,\ldots,p_k)\big) + \sum_{i=1}^{k} p_i S(\rho_i)$. For pseudo-density matrices (cf.~Remark~\ref{PDMs}), where negative eigenvalues render the standard von~Neumann entropy ill-defined, Ref.~\cite{Fullwood_Yang_2026} first established that the spectrum of every $N$-time PDM lies in $[-1,1]$. Since every PDM is Hermitian and of unit trace, its spectral decomposition $X = \sum_i \lambda_i \dyad{\psi_i}{\psi_i}$ takes the form of an affine combination of orthogonal pure states with weights $\lambda_i \in [-1,1]$ summing to unity. This observation motivates generalizing strong additivity to affine combinations with weights $(p_1,\ldots,p_k) \in [-1,1]^k$ summing to unity, which for an entropy functional $S$ yields the condition
\be\label{eq:pdm-sa}
S\!\left(\bigoplus_{i=1}^{k} p_i X_i\right) = S\big(\operatorname{diag}(p_1,\ldots,p_k)\big) + \sum_{i=1}^{k} p_i S(X_i)\, ,
\ee
where $\{X_1,\ldots,X_k\}$ is any finite collection of unit-trace Hermitian matrices. As shown in Ref.~\cite{Fullwood_Yang_2026}, unitary invariance $S(UXU^{\dag}) = S(X)$ and strong additivity~\eqref{eq:pdm-sa} over $[-1,1]$ isolate a unique extension of von~Neumann entropy to all unit-trace Hermitian matrices. Termed \emph{PDM entropy} in Ref.~\cite{Fullwood_Yang_2026}, this entropy functional is given by $S(X) = -\sum_i \lambda_i\log|\lambda_i| = -\mathrm{Tr}(X\log|X|)$.

Turning to the non-Hermitian setting of spacetime states, Proposition~\ref{prop:kd-properties}~(\ref{item:kd-marginals}) and~(\ref{item:kd-norm}) show that every $N$-time KD state over time $T_N$ and its Hermitian adjoint $\Sigma_N = T_N^\dagger$ are unit-trace contractions, satisfying $\mathrm{Tr}(T_N) = 1$ and $\|T_N\|_\infty \le 1$. In particular, all singular values of $T_N$ lie in $[0,1]$ and all eigenvalues of $T_N$ lie in the closed unit disk $\mathbb{D} = \{z \in \C : |z| \le 1\}$. One might therefore expect that imposing continuity alongside unitary invariance and strong additivity~\eqref{eq:pdm-sa} over affine combinations with weights $(p_1,\ldots,p_k) \in \mathbb{D}^k$ summing to unity would again single out a unique extension of von~Neumann entropy to unit-trace contractions. However, these assumptions are no longer sufficient once Hermiticity is dropped. Unlike PDMs, spacetime states are rarely normal, and are thus not unitarily diagonalizable in general (nor even necessarily diagonalizable). Consequently, unitary invariance no longer reduces the evaluation of an entropy functional to diagonal matrices. In fact, while the spectral functional $S_{\mathrm{spec}}(X) = -\sum_i \lambda_i\log|\lambda_i|$ (where the $\lambda_i$ are the eigenvalues of $X$ counted with algebraic multiplicity, often termed modified pseudo-entropy~\cite{Tu22,Salek_2014,Cipolloni_2023,Chen_2026}) and the operator-theoretic extension $S(X) = -\mathrm{Tr}(X\log\sqrt{X^\dagger X}) = -\mathrm{Tr}(X\log\sqrt{XX^\dagger})$ agree on all normal contractions, they do not coincide on general non-normal operators (including spacetime states). Nevertheless, both functionals reduce to von~Neumann entropy on density operators and to PDM entropy on Hermitian contractions, are continuous and unitarily invariant, and satisfy strong additivity over $\mathbb{D}$---as does every affine combination of the two.

While the spectral candidate $S_{\mathrm{spec}}(X) = -\sum_i \lambda_i\log|\lambda_i|$ may initially appear reasonable, it suffers from a fundamental structural limitation. Unlike the standard formulation of von~Neumann entropy, $S_{\mathrm{spec}}$ is defined solely in terms of eigenvalues rather than through the functional calculus. In quantum information theory, defining the von~Neumann entropy as $S(\rho) = -\mathrm{Tr}(\rho\log\rho)$ rather than the Shannon entropy of the spectrum of $\rho$ is far more than a notational convenience. The operator form identifies entropy directly as the expectation value of the Hermitian observable $K(\rho) = -\log\rho$, and it is precisely this algebraic structure that gives rise to the quantum relative entropy $D(\rho\|\sigma) = \mathrm{Tr}\big(\rho(\log\rho - \log\sigma)\big)$, where the misalignment between the eigenbases of $\rho$ and $\sigma$ carries essential physical content inaccessible to their spectra alone. A purely spectral functional such as $S_{\mathrm{spec}}$ supplies no analogous observable and therefore does not lend itself naturally to a relative formulation comparing distinct spacetime states. Retaining the principle that quantum entropy arises as the expectation value of a covariant operator associated with the state keeps the formalism anchored in the functional calculus and operator algebra, and as we show below, elevates this viewpoint into an axiom that isolates a unique spacetime entropy functional.

To make this framework precise, for each integer $n \ge 1$ let $M_n(\C)$ denote the algebra of $n \times n$ complex matrices, and let
\[
\mathcal{C}_n^1 = \big\{X \in M_n(\C) : \mathrm{Tr}(X) = 1 \text{ and } \|X\|_\infty \le 1\big\}\, ,
\qquad
\mathcal{C}^1 = \bigsqcup_{n=1}^{\infty} \mathcal{C}_n^1
\]
denote the space of finite-dimensional unit-trace contractions. We denote the unit circle by $\mathbb{T} = \{z \in \C : |z| = 1\}$, and for any matrix $X \in M_n(\C)$ we write $|X| = \sqrt{X^\dagger X}$ for its modulus. Given a positive semidefinite matrix $A \in M_n(\C)$, we define $\log A$ via the functional calculus applied to the function $t \mapsto \log t$ for $t > 0$ with the convention $\log 0 = 0$. Letting $X = W|X|$ be a polar decomposition of $X$ with $W \in M_n(\C)$ unitary, one has $X\log|X| = W g(|X|)$, where $g \colon [0,\infty) \to \R$ is the continuous function defined by $g(t) = t\log t$ for $t > 0$ and $g(0) = 0$. Since $\operatorname{im}(g(|X|)) \subseteq \operatorname{im}(|X|)$, the operator $X\log|X|$ is independent of the choice of unitary polar factor $W$. Moreover, the polar decomposition separates $X$ into its unitary factor $W$ and its positive semidefinite modulus $|X|$, suggesting that any operator $K(X)$ generalizing $-\log\rho$ should transform covariantly under unitary conjugation and depend solely on the modulus $|X|$. This motivates the following definition.

\bd[Polar expectation value]\label{def:mod-exp}
A functional $S \colon \mathcal{C}^1 \to \C$ is said to be a \define{polar expectation value} if for every integer $n \ge 1$ there exists a map $K_n \colon \mathcal{C}_n^1 \to M_n(\C)$ (written simply as $K$ when the dimension is clear from context) satisfying the following conditions for all $X, Y \in \mathcal{C}_n^1$.
\begin{enumerate}[(i)]
\item \label{item:mod-trace} \textbf{Expectation value form.} $S(X) = \mathrm{Tr}\big(X K(X)\big)$.
\item \label{item:mod-equiv} \textbf{Unitary equivariance.} $K(V X V^\dagger) = V K(X) V^\dagger$ for every unitary $V \in M_n(\C)$.
\item \label{item:mod-abs} \textbf{Modulus dependence.} $K(X) = K(Y)$ whenever $|X| = |Y|$.
\end{enumerate}
\ed

Note that conditions~(\ref{item:mod-trace}) and~(\ref{item:mod-equiv}) in Definition~\ref{def:mod-exp} automatically imply unitary invariance of $S$, since for every $X \in \mathcal{C}_n^1$ and every unitary $V \in M_n(\C)$,
\[
S(V X V^\dagger) = \mathrm{Tr}\big(V X V^\dagger K(V X V^\dagger)\big) = \mathrm{Tr}\big(V X K(X) V^\dagger\big) = \mathrm{Tr}\big(X K(X)\big) = S(X)\, .
\]
Moreover, no continuity assumption is imposed on the operator-valued map $K$ itself, which is essential as the logarithmic operator $-\log|X|$ is discontinuous as a singular value approaches zero. We now state our main characterization theorem, which uses four simple axioms to single out a unique entropy functional that we refer to as \emph{spacetime entropy}.

\bt[Axiomatic characterization of spacetime entropy]\label{thm:spacetime-entropy}
Suppose $S \colon \mathcal{C}^1 \to \C$ is a functional on the space of finite-dimensional unit-trace contractions satisfying the following conditions.
\begin{enumerate}[(1)]
\item \label{item:ax-vn} \textbf{Reduction to von~Neumann entropy.} $S(\rho) = -\mathrm{Tr}(\rho\log\rho)$ for every positive semidefinite element $\rho \in \mathcal{C}^1$.
\item \label{item:ax-cont} \textbf{Continuity.} The restriction of $S$ to $\mathcal{C}_n^1$ is continuous for every integer $n \ge 1$.
\item \label{item:ax-sa} \textbf{Strong additivity over $\mathbb{D}$.} Whenever $X_i \in \mathcal{C}_{n_i}^1$ ($i = 1,\ldots,k$) and $(p_1,\ldots,p_k) \in \mathbb{D}^k$ satisfy $\sum_{i=1}^{k} p_i = 1$,
\be\label{eq:sa-disk}
S\!\left(\bigoplus_{i=1}^{k} p_i X_i\right) = S\big(\operatorname{diag}(p_1,\ldots,p_k)\big) + \sum_{i=1}^{k} p_i S(X_i)\, .
\ee
\item \label{item:ax-mod} \textbf{Polar expectation value.} $S$ is a polar expectation value in the sense of Definition~\ref{def:mod-exp}.
\end{enumerate}
Then for every $X \in \mathcal{C}^1$,
\be\label{eq:spacetime-entropy-formula}
S(X) = -\mathrm{Tr}\big(X\log|X|\big) = -\mathrm{Tr}\big(X\log\sqrt{X^\dagger X}\big) = -\mathrm{Tr}\big(X\log\sqrt{XX^\dagger}\big)\, .
\ee
Conversely, the functional $S \colon \mathcal{C}^1 \to \C$ defined by~\eqref{eq:spacetime-entropy-formula} satisfies conditions~(\ref{item:ax-vn})--(\ref{item:ax-mod}).
\et

Before proving Theorem~\ref{thm:spacetime-entropy}, we state two preliminary lemmas whose proofs are deferred to Appendix~\ref{app:lemmas}. To fix notation for the first lemma, for each integer $n \ge 1$ let
\[
\mathcal{V}_n(\C) = \left\{(z_1,\ldots,z_n) \in \C^n : \sum_{j=1}^{n} z_j = 1\right\}\, ,
\qquad
\Q_n(\mathbb{D}) = \mathcal{V}_n(\C) \cap \mathbb{D}^n\, ,
\qquad
\Delta_n = \mathcal{V}_n(\C) \cap [0,1]^n\, ,
\]
and set $\mathcal{V}(\C) = \bigcup_{n=1}^{\infty} \mathcal{V}_n(\C)$, $\Q(\mathbb{D}) = \bigcup_{n=1}^{\infty} \Q_n(\mathbb{D})$, and $\Delta = \bigcup_{n=1}^{\infty} \Delta_n$. Given $p = (p_1,\ldots,p_k) \in \mathcal{V}_k(\C)$ and $q^{(i)} = (q_{i1},\ldots,q_{im_i}) \in \mathcal{V}_{m_i}(\C)$ ($i=1,\ldots,k$), we denote their weighted concatenation in $\mathcal{V}_{m_1+\cdots+m_k}(\C)$ by
\[
\bigoplus_{i=1}^{k} p_i q^{(i)} = \big(p_1 q_{11},\ldots,p_1 q_{1m_1},\ldots,p_k q_{k1},\ldots,p_k q_{km_k}\big)\, .
\]

\blem\label{lem:complex-strong-add}
Let $h \colon \Q(\mathbb{D}) \to \C$ be a function that is permutation invariant and continuous on each $\Q_n(\mathbb{D})$, satisfies $h(p_1,\ldots,p_n) = -\sum_{j=1}^{n} p_j\log p_j$ for all $(p_1,\ldots,p_n) \in \Delta$, and obeys the strong additivity relation
\be\label{eq:sa-vec}
h\!\left(\bigoplus_{i=1}^{k} p_i q^{(i)}\right) = h(p_1,\ldots,p_k) + \sum_{i=1}^{k} p_i h\big(q^{(i)}\big)
\ee
for all $p = (p_1,\ldots,p_k) \in \Q_k(\mathbb{D})$ and $q^{(i)} \in \Q_{m_i}(\mathbb{D})$. Then
\be\label{eq:h-complex-formula}
h(z_1,\ldots,z_n) = -\sum_{j=1}^{n} z_j \log|z_j| \qquad \forall (z_1,\ldots,z_n) \in \Q(\mathbb{D})\, ,
\ee
with the convention $0\log 0 = 0$.
\elem

\blem\label{lem:unit-polygon}
Let $m \ge 1$ be an integer, and let $c_1,\ldots,c_m \in (0,1]$ satisfy $\sum_{t=1}^{m} c_t \ge 1$. Then there exist phases $z_1,\ldots,z_m \in \mathbb{T}$ such that $\sum_{t=1}^{m} c_t z_t = 1$.
\elem

With Lemmas~\ref{lem:complex-strong-add} and~\ref{lem:unit-polygon} in hand, we now prove Theorem~\ref{thm:spacetime-entropy}.

\bprf[Proof of Theorem~\ref{thm:spacetime-entropy}]
First suppose that $S \colon \mathcal{C}^1 \to \C$ satisfies conditions~(\ref{item:ax-vn})--(\ref{item:ax-mod}), with associated operator-valued maps $K \colon \mathcal{C}_n^1 \to M_n(\C)$ from Definition~\ref{def:mod-exp}. We first note that the final equality in~\eqref{eq:spacetime-entropy-formula} holds identically for every $X \in M_n(\C)$. Indeed, writing a polar decomposition $X = W P$ with $W \in M_n(\C)$ unitary and $P = |X| = \sqrt{X^\dagger X}$, one has $\sqrt{XX^\dagger} = W P W^\dagger$ and hence $\log\sqrt{XX^\dagger} = W(\log P)W^\dagger$. Using cyclicity of the trace and the fact that $P$ commutes with $\log P$ yields
\begin{align*}
\mathrm{Tr}\big(X\log\sqrt{XX^\dagger}\big) &= \mathrm{Tr}\big(W P W(\log P) W^\dagger\big) = \mathrm{Tr}\big(P W \log P\big) \\
& = \mathrm{Tr}\big((\log P) P W\big) = \mathrm{Tr}\big(W P \log P\big) = \mathrm{Tr}\big(X\log|X|\big)\, .
\end{align*}

To show that $S(X) = -\mathrm{Tr}(X\log|X|)$ for all $X \in \mathcal{C}^1$, first define $h \colon \Q(\mathbb{D}) \to \C$ by $h(z_1,\ldots,z_n) = S\big(\operatorname{diag}(z_1,\ldots,z_n)\big)$. By unitary invariance of $S$ (implied by condition~(\ref{item:ax-mod})), $h$ is permutation invariant. By conditions~(\ref{item:ax-vn})--(\ref{item:ax-sa}), $h$ is continuous on each $\Q_n(\mathbb{D})$, reduces to Shannon entropy on $\Delta$, and satisfies~\eqref{eq:sa-vec}. Lemma~\ref{lem:complex-strong-add} therefore yields
\be\label{eq:S-diag}
S\big(\operatorname{diag}(z_1,\ldots,z_n)\big) = -\sum_{j=1}^{n} z_j \log|z_j| \qquad \forall (z_1,\ldots,z_n) \in \Q_n(\mathbb{D})\, .
\ee

Now fix an arbitrary $X \in \mathcal{C}_n^1$, and let $X = W P$ be a polar decomposition with $W \in M_n(\C)$ unitary and $P = |X| \ge 0$. By unitary invariance of $S$ and the fact that $-\mathrm{Tr}(X\log|X|)$ is unitarily invariant, we may conjugate $X$ by a unitary matrix that diagonalizes $P$ and assume without loss of generality that
\[
P = \operatorname{diag}(\sigma_1,\ldots,\sigma_r, 0,\ldots,0)\, ,
\]
where $r = \operatorname{rank}(X) \ge 1$ and $1 \ge \sigma_1 \ge \sigma_2 \ge \cdots \ge \sigma_r > 0$. For each $p \in \{1,\ldots,n\}$, the diagonal reflection $V_p = \operatorname{diag}(1,\ldots,1,-1,1,\ldots,1)$ (with $-1$ in the $p$-th position) is unitary and commutes with $P$, so $|V_p X V_p^\dagger| = V_p P V_p^\dagger = P = |X|$. By conditions~(\ref{item:mod-equiv}) and~(\ref{item:mod-abs}) of Definition~\ref{def:mod-exp}, we obtain
\[
V_p K(X) V_p^\dagger = K(V_p X V_p^\dagger) = K(X)\, .
\]
Since conjugating by $V_p$ negates the $(p,q)$ entry for all $q \ne p$, the matrix $K(X)$ is diagonal, say $K(X) = \operatorname{diag}(k_1,\ldots,k_n)$. Writing $W_{jj}$ for the diagonal entries of $W$, we then have
\be\label{eq:S-X-diagK}
S(X) = \mathrm{Tr}\big(X K(X)\big) = \mathrm{Tr}\big(W P K(X)\big) = \sum_{j=1}^{r} W_{jj} \sigma_j k_j\, .
\ee

Since $W$ is unitary, $|W_{jj}| \le 1$ for every $j$, which yields
\[
1 = |\mathrm{Tr}(X)| = \left|\sum_{j=1}^{r} W_{jj}\sigma_j\right| \le \sum_{j=1}^{r} |W_{jj}|\sigma_j \le \sum_{j=1}^{r} \sigma_j\, .
\]
If $\sum_{j=1}^{r} \sigma_j = 1$, then taking real parts of $\sum_{j=1}^{r} W_{jj}\sigma_j = 1$ gives $\sum_{j=1}^{r} (1 - \operatorname{Re} W_{jj})\sigma_j = 0$, which forces $W_{jj} = 1$ for all $j \in \{1,\ldots,r\}$. As each column of $W$ has unit Euclidean norm, all off-diagonal entries in the first $r$ columns of $W$ vanish, so $X = W P = P$ is positive semidefinite. Condition~(\ref{item:ax-vn}) then immediately gives $S(X) = -\mathrm{Tr}(P\log P) = -\mathrm{Tr}(X\log|X|)$.

Assume now that $\sum_{j=1}^{r} \sigma_j > 1$. Since $\sigma_1 \le 1$, this forces $r \ge 2$. Define
\[
\mathcal{M} = \left\{z = (z_1,\ldots,z_r) \in \mathbb{T}^r : \sum_{j=1}^{r} \sigma_j z_j = 1\right\}\, .
\]
For every $z \in \mathcal{M}$, the diagonal matrix $Y_z = \operatorname{diag}(z_1\sigma_1,\ldots,z_r\sigma_r,0,\ldots,0)$ satisfies $\mathrm{Tr}(Y_z) = 1$, $\|Y_z\|_\infty = \sigma_1 \le 1$, and $|Y_z| = P = |X|$. Consequently, $Y_z \in \mathcal{C}_n^1$ and $K(Y_z) = K(X)$ by modulus dependence. Equating $S(Y_z) = \mathrm{Tr}(Y_z K(Y_z)) = \sum_{j=1}^{r} z_j \sigma_j k_j$ with~\eqref{eq:S-diag} gives
\be\label{eq:a-vanish}
\sum_{j=1}^{r} a_j \sigma_j z_j = 0 \qquad \forall z \in \mathcal{M}\, ,
\ee
where $a_j = k_j + \log\sigma_j$ for $j \in \{1,\ldots,r\}$.

Fix any $k \in \{2,\ldots,r\}$. We claim that there exist phases $(z_\ell)_{\ell \notin \{1,k\}}$ in $\mathbb{T}$ (an empty tuple when $r = 2$) such that $w = 1 - \sum_{\ell \notin \{1,k\}} \sigma_\ell z_\ell$ satisfies the strict inequalities
\be\label{eq:strict-tri}
\sigma_1 - \sigma_k < |w| < \sigma_1 + \sigma_k\, .
\ee
When $r = 2$, we have $w = 1$, which satisfies~\eqref{eq:strict-tri} directly since $\sigma_1 \le 1$, $\sigma_2 > 0$, and $\sigma_1 + \sigma_2 > 1$. When $r \ge 3$, set $S' = \sum_{\ell \notin \{1,k\}} \sigma_\ell > 0$ and consider the continuous function $F \colon \mathbb{T}^{r-2} \to [0,\infty)$ given by $F\big((z_\ell)_{\ell \notin \{1,k\}}\big) = \big|1 - \sum_{\ell \notin \{1,k\}} \sigma_\ell z_\ell\big|$. Choosing $z_\ell = -1$ for all $\ell$ yields $F = 1 + S' > 1 > \sigma_1 - \sigma_k$. On the other hand, if $S' \le 1$, then choosing $z_\ell = 1$ for all $\ell$ yields $F = 1 - S' < \sigma_1 + \sigma_k$, while if $S' > 1$, then Lemma~\ref{lem:unit-polygon} yields $(z_\ell)_{\ell \notin \{1,k\}} \in \mathbb{T}^{r-2}$ with $\sum_{\ell \notin \{1,k\}} \sigma_\ell z_\ell = 1$, giving $F = 0 < \sigma_1 + \sigma_k$. Since $\mathbb{T}^{r-2}$ is connected, the image $F(\mathbb{T}^{r-2})$ is a closed interval containing a value strictly less than $\sigma_1 + \sigma_k$ and a value strictly greater than $\sigma_1 - \sigma_k$, so it intersects the open interval $(\sigma_1 - \sigma_k, \sigma_1 + \sigma_k)$, establishing~\eqref{eq:strict-tri}.

Since $|w| > \sigma_1 - \sigma_k \ge 0$, we may write $w = |w| u_0$ with $u_0 \in \mathbb{T}$. Define
\[
\gamma = \frac{|w|^2 + \sigma_1^2 - \sigma_k^2}{2|w|\sigma_1}\, .
\]
From~\eqref{eq:strict-tri} and $\sigma_1 \ge \sigma_k$ we obtain
\[
1 - \gamma = \frac{(\sigma_k - |w| + \sigma_1)(\sigma_k + |w| - \sigma_1)}{2|w|\sigma_1} > 0\, ,
\qquad
1 + \gamma = \frac{(|w| + \sigma_1 - \sigma_k)(|w| + \sigma_1 + \sigma_k)}{2|w|\sigma_1} > 0\, ,
\]
so $|\gamma| < 1$. Choosing $\theta \in (0,\pi)$ with $\cos\theta = \gamma$ (whence $\sin\theta > 0$), set $\zeta_k = \sigma_k^{-1}(|w| - \sigma_1 e^{i\theta})$. A direct computation gives $|\zeta_k|^2 = \sigma_k^{-2}(|w|^2 + \sigma_1^2 - 2|w|\sigma_1\cos\theta) = 1$, so $\zeta_k \in \mathbb{T}$. Setting $z_1^\pm = u_0 e^{\pm i\theta} \in \mathbb{T}$ and $z_k^+ = u_0 \zeta_k$, $z_k^- = u_0 \overline{\zeta_k} \in \mathbb{T}$, we have
\[
\sigma_1 z_1^\pm + \sigma_k z_k^\pm = u_0\big(\sigma_1 e^{\pm i\theta} + (|w| - \sigma_1 e^{\pm i\theta})\big) = w = 1 - \sum_{\ell \notin \{1,k\}} \sigma_\ell z_\ell\, .
\]
Hence the two vectors $z^+, z^- \in \mathbb{T}^r$ formed by combining $(z_1^\pm, z_k^\pm)$ with the fixed phases $(z_\ell)_{\ell \notin \{1,k\}}$ both belong to $\mathcal{M}$. Subtracting $\sigma_1 z_1^- + \sigma_k z_k^- = w$ from $\sigma_1 z_1^+ + \sigma_k z_k^+ = w$ gives $\sigma_k(z_k^+ - z_k^-) = -\sigma_1(z_1^+ - z_1^-)$. Subtracting~\eqref{eq:a-vanish} evaluated at $z^-$ from~\eqref{eq:a-vanish} evaluated at $z^+$ then yields
\[
0 = a_1 \sigma_1 (z_1^+ - z_1^-) + a_k \sigma_k (z_k^+ - z_k^-) = (a_1 - a_k)\sigma_1 (z_1^+ - z_1^-)\, .
\]
Since $z_1^+ - z_1^- = 2 i u_0 \sin\theta \ne 0$ and $\sigma_1 > 0$, we conclude that $a_k = a_1$. As $k \in \{2,\ldots,r\}$ was arbitrary, $a_1 = \cdots = a_r = a$ for a single constant $a \in \C$. Substituting this into~\eqref{eq:a-vanish} at $z^+ \in \mathcal{M}$ gives $0 = a \sum_{j=1}^{r} \sigma_j z_j^+ = a$, so $k_j = -\log\sigma_j$ for every $j \in \{1,\ldots,r\}$. By~\eqref{eq:S-X-diagK}, we obtain
\[
S(X) = -\sum_{j=1}^{r} W_{jj}\sigma_j\log\sigma_j = -\mathrm{Tr}\big(W P \log P\big) = -\mathrm{Tr}\big(X\log|X|\big)\, .
\]

Conversely, let $S \colon \mathcal{C}^1 \to \C$ be defined by $S(X) = -\mathrm{Tr}(X\log|X|)$. Condition~(\ref{item:ax-vn}) is immediate, since $|\rho| = \rho$ for every positive semidefinite $\rho \in \mathcal{C}^1$. To verify condition~(\ref{item:ax-cont}), let $(X_m)_{m=1}^{\infty}$ be a sequence in $\mathcal{C}_n^1$ converging to $X \in \mathcal{C}_n^1$, and choose polar decompositions $X_m = W_m |X_m|$ with $W_m \in M_n(\C)$ unitary. By compactness of the unitary group in $M_n(\C)$, every subsequence of $(W_m)$ admits a further subsequence $(W_{m_j})$ converging to a unitary $W_*$. Since the map $Y \mapsto |Y|$ and the function $g(t) = t\log t$ ($g(0) = 0$) are continuous, we have $X = \lim_j W_{m_j}|X_{m_j}| = W_* |X|$ and $\lim_j X_{m_j}\log|X_{m_j}| = \lim_j W_{m_j} g(|X_{m_j}|) = W_* g(|X|) = X\log|X|$. Since this limit is independent of the chosen subsequence, $X_m\log|X_m| \to X\log|X|$, and taking traces shows that $S$ is continuous on $\mathcal{C}_n^1$.

To verify condition~(\ref{item:ax-sa}), let $X_i \in \mathcal{C}_{n_i}^1$ ($i=1,\ldots,k$) and $(p_1,\ldots,p_k) \in \Q_k(\mathbb{D})$, and set $X = \bigoplus_{i=1}^{k} p_i X_i \in \mathcal{C}_{n_1+\cdots+n_k}^1$. Then $|X| = \bigoplus_{i=1}^{k} |p_i|\,|X_i|$. For each $i$ with $p_i = 0$, the $i$-th block contributes $0 = -p_i\log|p_i| + p_i S(X_i)$. For each $i$ with $p_i \ne 0$, let $\Pi_i$ denote the orthogonal projector onto $\operatorname{im}(|X_i|)$. Under our convention $\log 0 = 0$, one has $\log(|p_i|\,|X_i|) = (\log|p_i|)\Pi_i + \log|X_i|$. Since $X_i \Pi_i = X_i$ and $\mathrm{Tr}(X_i) = 1$, it follows that
\[
-\mathrm{Tr}\big(p_i X_i \log(|p_i|\,|X_i|)\big) = -p_i\log|p_i| - p_i\mathrm{Tr}\big(X_i\log|X_i|\big) = -p_i\log|p_i| + p_i S(X_i)\, .
\]
Summing over $i = 1,\ldots,k$ yields~\eqref{eq:sa-disk}. Finally, setting $K(X) = -\log|X|$ for $X \in \mathcal{C}_n^1$ directly satisfies conditions~(\ref{item:mod-trace})--(\ref{item:mod-abs}) of Definition~\ref{def:mod-exp}, since $|V X V^\dagger| = V|X|V^\dagger$ implies $K(V X V^\dagger) = -\log(V|X|V^\dagger) = V K(X) V^\dagger$ for every unitary $V \in M_n(\C)$. Thus $S$ satisfies condition~(\ref{item:ax-mod}), completing the proof.
\eprf

\br[Left-modulus symmetry]\label{rem:entropy-adjoint}
In Definition~\ref{def:mod-exp}, replacing the right modulus $|X| = \sqrt{X^\dagger X}$ in condition~(\ref{item:mod-abs}) by the left modulus $\sqrt{XX^\dagger}$ yields an equivalent characterization of the same functional~\eqref{eq:spacetime-entropy-formula}, as reflected by the identity $\mathrm{Tr}\big(X\log\sqrt{XX^\dagger}\big) = \mathrm{Tr}\big(X\log\sqrt{X^\dagger X}\big)$ established in the proof of Theorem~\ref{thm:spacetime-entropy}.
\er

We now establish several basic properties of spacetime entropy that mirror fundamental properties of the von~Neumann entropy.

\bn[Basic properties of spacetime entropy]\label{prop:spacetime-entropy-properties}
Let $S \colon \mathcal{C}^1 \to \C$ denote the spacetime entropy functional $S(X) = -\mathrm{Tr}(X\log|X|)$ characterized in Theorem~\ref{thm:spacetime-entropy}. Then $S$ satisfies the following properties.
\begin{enumerate}[(i)]
\item \label{item:se-tensor} \textbf{Additivity over tensor products.} For all $X, Y \in \mathcal{C}^1$, one has $X \otimes Y \in \mathcal{C}^1$ and
\be\label{eq:se-tensor}
S(X \otimes Y) = S(X) + S(Y)\, .
\ee
In particular, for any two $N$-time quantum dynamics $(\rho_1,\E_1,\ldots,\E_{N-1})$ and $(\sigma_1,\F_1,\ldots,\F_{N-1})$ evolving in parallel,
\be\label{eq:se-parallel}
S\big(T_N(\rho_1 \otimes \sigma_1,\E_1 \otimes \F_1,\ldots,\E_{N-1} \otimes \F_{N-1})\big) = S\big(T_N(\rho_1,\E_1,\ldots,\E_{N-1})\big) + S\big(T_N(\sigma_1,\F_1,\ldots,\F_{N-1})\big)\, .
\ee

\item \label{item:se-sym} \textbf{Symmetries under adjoint, transpose, and conjugation.} For every $X \in \mathcal{C}_n^1$, the operators $X^\dagger$, $X^T$, and $\overline{X}$ (with transpose and entrywise complex conjugation taken in any fixed orthonormal basis) belong to $\mathcal{C}_n^1$ and satisfy
\be\label{eq:se-sym}
S(X^\dagger) = \overline{S(X)}\, ,
\qquad
S(X^T) = S(X)\, ,
\qquad
S(\overline{X}) = \overline{S(X)}\, .
\ee
In particular, for any $N$-time quantum dynamics with KD state over time $T_N$ and spacetime density matrix $\Sigma_N$, one has $S(\Sigma_N) = \overline{S(T_N)}$.

\item \label{item:se-bound} \textbf{Singular-value bound.} For every $X \in \mathcal{C}_n^1$ with singular values $\sigma_1,\ldots,\sigma_n \in [0,1]$,
\be\label{eq:se-bound}
|S(X)| \le -\mathrm{Tr}\big(|X|\log|X|\big) = -\sum_{j=1}^{n} \sigma_j \log\sigma_j\, ,
\ee
with equality whenever $X$ is positive semidefinite.

\item \label{item:se-isom} \textbf{Invariance under isometric and unitary dynamics.} Let $(\rho_1,\E_1,\ldots,\E_{N-1})$ be an $N$-time quantum dynamics in which each channel is isometric, meaning that $\E_k(A) = V_k A V_k^\dagger$ for an isometry $V_k \colon \H_k \to \H_{k+1}$ ($V_k^\dagger V_k = \one_k$) for all $k \in \{1,\ldots,N-1\}$. Then
\be\label{eq:se-isom}
S(T_N) = S(\Sigma_N) = S(\rho_1)\, .
\ee
In particular, for time-homogeneous unitary evolution, the $N$-time spacetime state $R_N$ of Definition~\ref{def:spacetime} satisfies $S(R_N) = S(\rho_1)$, which vanishes whenever $\rho_1$ is pure.
\end{enumerate}
\en

\bprf
\noindent\textbf{Item~(\ref{item:se-tensor}).} Let $X \in \mathcal{C}_m^1$ and $Y \in \mathcal{C}_n^1$. Then $\mathrm{Tr}(X \otimes Y) = \mathrm{Tr}(X)\mathrm{Tr}(Y) = 1$ and $\|X \otimes Y\|_\infty = \|X\|_\infty \|Y\|_\infty \le 1$, so $X \otimes Y \in \mathcal{C}_{mn}^1$. Since $|X| \otimes |Y| \ge 0$ and $(|X| \otimes |Y|)^2 = (X^\dagger X) \otimes (Y^\dagger Y) = (X \otimes Y)^\dagger (X \otimes Y)$, uniqueness of the positive semidefinite square root yields $|X \otimes Y| = |X| \otimes |Y|$. Let $\Pi_X \in M_m(\C)$ and $\Pi_Y \in M_n(\C)$ denote the orthogonal projectors onto $\operatorname{im}(|X|)$ and $\operatorname{im}(|Y|)$, respectively. Choosing orthonormal eigenbases $\{\ket{u_a}\}$ of $|X|$ with eigenvalues $\sigma_a \ge 0$ and $\{\ket{v_b}\}$ of $|Y|$ with eigenvalues $\tau_b \ge 0$, every product basis vector $\ket{u_a} \otimes \ket{v_b}$ is an eigenvector of $|X| \otimes |Y|$ with eigenvalue $\sigma_a \tau_b$. Under the convention $\log 0 = 0$, the eigenvalue of $\log(|X| \otimes |Y|)$ on $\ket{u_a} \otimes \ket{v_b}$ equals $\log\sigma_a + \log\tau_b$ when $\sigma_a > 0$ and $\tau_b > 0$, and equals $0$ whenever $\sigma_a = 0$ or $\tau_b = 0$. Comparing with the action of $(\log|X|) \otimes \Pi_Y + \Pi_X \otimes (\log|Y|)$ on $\ket{u_a} \otimes \ket{v_b}$ gives the operator identity
\[
\log|X \otimes Y| = (\log|X|) \otimes \Pi_Y + \Pi_X \otimes (\log|Y|)\, .
\]
Writing polar decompositions $X = W_X |X|$ and $Y = W_Y |Y|$ with $W_X, W_Y$ unitary, the relations $|X|\Pi_X = |X|$ and $|Y|\Pi_Y = |Y|$ imply $X\Pi_X = X$ and $Y\Pi_Y = Y$. Therefore,
\begin{align*}
S(X \otimes Y)
&= -\mathrm{Tr}\Big[(X \otimes Y)\big((\log|X|) \otimes \Pi_Y + \Pi_X \otimes (\log|Y|)\big)\Big] \\
&= -\mathrm{Tr}\big(X\log|X|\big)\mathrm{Tr}(Y\Pi_Y) - \mathrm{Tr}(X\Pi_X)\mathrm{Tr}\big(Y\log|Y|\big) \\
&= -\mathrm{Tr}\big(X\log|X|\big)\mathrm{Tr}(Y) - \mathrm{Tr}(X)\mathrm{Tr}\big(Y\log|Y|\big) \\
&= S(X) + S(Y)\, ,
\end{align*}
establishing~\eqref{eq:se-tensor}. Combining~\eqref{eq:se-tensor} with unitary invariance of $S$ and Proposition~\ref{prop:kd-properties}~(\ref{item:kd-parallel}) immediately yields~\eqref{eq:se-parallel}.

\medskip
\noindent\textbf{Item~(\ref{item:se-sym}).} Let $X \in \mathcal{C}_n^1$. Clearly $\mathrm{Tr}(X^\dagger) = \overline{\mathrm{Tr}(X)} = 1$ and $\|X^\dagger\|_\infty = \|X\|_\infty \le 1$, so $X^\dagger \in \mathcal{C}_n^1$. Applying the identity $\mathrm{Tr}\big(Y\log\sqrt{Y^\dagger Y}\big) = \mathrm{Tr}\big(Y\log\sqrt{YY^\dagger}\big)$ from~\eqref{eq:spacetime-entropy-formula} to $Y = X^\dagger$ and using the Hermiticity of $|X| = \sqrt{X^\dagger X}$ yields
\[
S(X^\dagger) = -\mathrm{Tr}\big(X^\dagger\log\sqrt{XX^\dagger}\big) = -\mathrm{Tr}\big(X^\dagger\log|X|\big) = \overline{-\mathrm{Tr}\big((\log|X|)X\big)} = \overline{-\mathrm{Tr}\big(X\log|X|\big)} = \overline{S(X)}\, .
\]
By Theorem~\ref{thm:sdm-equiv}, it follows immediately that $S(\Sigma_N) = S(T_N^\dagger) = \overline{S(T_N)}$. Next, fix an orthonormal basis so that transpose and entrywise complex conjugation are defined on $M_n(\C)$. Then $\mathrm{Tr}(X^T) = \mathrm{Tr}(X) = 1$, and $(X^T)^\dagger X^T = \overline{X}X^T = (XX^\dagger)^T$ has the same eigenvalues as $XX^\dagger$, whence $\|X^T\|_\infty = \|X\|_\infty \le 1$ and $X^T \in \mathcal{C}_n^1$. For any positive semidefinite matrix $A = \sum_j \lambda_j \dyad{\psi_j}{\psi_j}$, the transposed spectral decomposition $A^T = \sum_j \lambda_j \dyad{\overline{\psi_j}}{\overline{\psi_j}}$ shows that $A^T \ge 0$ and $f(A^T) = f(A)^T$ for every function $f \colon [0,\infty) \to \R$. Applying this to $A = \sqrt{XX^\dagger}$ gives $|X^T| = \sqrt{(XX^\dagger)^T} = (\sqrt{XX^\dagger})^T$ and hence $\log|X^T| = (\log\sqrt{XX^\dagger})^T$. Using $\mathrm{Tr}(M^T) = \mathrm{Tr}(M)$ and cyclicity of the trace together with~\eqref{eq:spacetime-entropy-formula}, we obtain
\[
S(X^T) = -\mathrm{Tr}\Big(X^T\big(\log\sqrt{XX^\dagger}\big)^T\Big) = -\mathrm{Tr}\Big(\big((\log\sqrt{XX^\dagger})X\big)^T\Big) = -\mathrm{Tr}\big(X\log\sqrt{XX^\dagger}\big) = S(X)\, .
\]
Finally, since $\overline{X} = (X^\dagger)^T$, we have $\overline{X} \in \mathcal{C}_n^1$ and $S(\overline{X}) = S\big((X^\dagger)^T\big) = S(X^\dagger) = \overline{S(X)}$, establishing~\eqref{eq:se-sym}.

\medskip
\noindent\textbf{Item~(\ref{item:se-bound}).} Let $X \in \mathcal{C}_n^1$ have polar decomposition $X = W|X|$ with $W \in M_n(\C)$ unitary, and choose an orthonormal eigenbasis $\{\ket{e_j}\}_{j=1}^{n}$ of $|X|$ with corresponding singular values $\sigma_1,\ldots,\sigma_n \in [0,1]$. Writing $W_{jj} = \bra{e_j}W\ket{e_j}$, we have $|W_{jj}| \le 1$ and $-\sigma_j\log\sigma_j \ge 0$ for every $j \in \{1,\ldots,n\}$. Applying the triangle inequality therefore yields
\[
|S(X)| = \left|-\sum_{j=1}^{n} W_{jj}\sigma_j\log\sigma_j\right| \le \sum_{j=1}^{n} |W_{jj}|(-\sigma_j\log\sigma_j) \le -\sum_{j=1}^{n} \sigma_j\log\sigma_j = -\mathrm{Tr}\big(|X|\log|X|\big)\, .
\]
When $X$ is positive semidefinite, $|X| = X$ and $S(X) = -\mathrm{Tr}(X\log X) \ge 0$, so equality holds in~\eqref{eq:se-bound}.

\medskip
\noindent\textbf{Item~(\ref{item:se-isom}).} For each $k \in \{1,\ldots,N-1\}$, let $P_{k+1} = V_k V_k^\dagger \in \Lin(\H_{k+1})$ denote the orthogonal projector onto $\operatorname{im}(V_k)$. Choosing an orthonormal basis $\{\ket{i}\}$ of $\H_k$ and using the Hermiticity of $\J[\E_k]$ together with $V_k^\dagger V_k = \one_k$, we compute
\begin{align*}
\J[\E_k]\J[\E_k]^\dagger
&= \sum_{i,j,a,b} \dyad{i}{j}\dyad{a}{b} \otimes V_k\dyad{j}{i}V_k^\dagger V_k\dyad{b}{a}V_k^\dagger \\
&= \sum_{i,j,a,b} \delta_{ja}\dyad{i}{b} \otimes \delta_{ib} V_k\dyad{j}{a}V_k^\dagger \\
&= \sum_{i,j} \dyad{i}{i} \otimes V_k\dyad{j}{j}V_k^\dagger
= \one_k \otimes V_k \one_k V_k^\dagger = \one_k \otimes P_{k+1}\, .
\end{align*}
We now prove by induction on $N > 1$ that
\be\label{eq:isom-TTdag}
T_N T_N^\dagger = \rho_1^2 \otimes P_2 \otimes \cdots \otimes P_N\, .
\ee
For $N = 2$, we have $T_2 = (\rho_1 \otimes \one_2)\J[\E_1]$, so
\[
T_2 T_2^\dagger = (\rho_1 \otimes \one_2)\J[\E_1]\J[\E_1]^\dagger(\rho_1 \otimes \one_2) = (\rho_1 \otimes \one_2)(\one_1 \otimes P_2)(\rho_1 \otimes \one_2) = \rho_1^2 \otimes P_2\, .
\]
Assuming~\eqref{eq:isom-TTdag} holds for a given $N > 1$, writing $T_{N+1} = (T_N \otimes \one_{N+1})\J_{N,N+1}[\E_N]$ and using $\J_{N,N+1}[\E_N]\J_{N,N+1}[\E_N]^\dagger = \one_{1\cdots N} \otimes P_{N+1}$ gives
\[
T_{N+1}T_{N+1}^\dagger = (T_N \otimes \one_{N+1})(\one_{1\cdots N} \otimes P_{N+1})(T_N^\dagger \otimes \one_{N+1}) = (T_N T_N^\dagger) \otimes P_{N+1} = \rho_1^2 \otimes P_2 \otimes \cdots \otimes P_{N+1}\, ,
\]
completing the induction. Since $\rho_1 \ge 0$ and each $P_k$ is an orthogonal projector ($P_k \ge 0$ and $P_k^2 = P_k$), taking the positive semidefinite square root of~\eqref{eq:isom-TTdag} yields $\sqrt{T_N T_N^\dagger} = \rho_1 \otimes P_2 \otimes \cdots \otimes P_N$. Evaluating $\log$ in a product eigenbasis of $\rho_1, P_2,\ldots,P_N$ under the convention $\log 0 = 0$ then gives $\log\sqrt{T_N T_N^\dagger} = (\log\rho_1) \otimes P_2 \otimes \cdots \otimes P_N$, as every eigenvalue of each $P_k$ lies in $\{0,1\}$. Applying the left-modulus formula in~\eqref{eq:spacetime-entropy-formula} together with the Kirkwood-Dirac correlator identity~\eqref{eq:kd-correlator} yields
\begin{align*}
S(T_N) &= -\mathrm{Tr}\big(T_N\log\sqrt{T_N T_N^\dagger}\big) = -\mathrm{Tr}\Big[T_N\big((\log\rho_1) \otimes P_2 \otimes \cdots \otimes P_N\big)\Big] \\
& = -\mathrm{Tr}\Big(P_N \E_{N-1}\big(\cdots P_2 \E_1((\log\rho_1)\rho_1)\cdots\big)\Big)\, .
\end{align*}
For every $k \in \{1,\ldots,N-1\}$ and every operator $A \in \Lin(\H_k)$, one has $P_{k+1}\E_k(A) = V_k V_k^\dagger V_k A V_k^\dagger = V_k A V_k^\dagger = \E_k(A)$, so each factor $P_{k+1}$ is absorbed into $\E_k$. Repeatedly invoking the trace-preserving property of $\E_1,\ldots,\E_{N-1}$ therefore yields
\[
S(T_N) = -\mathrm{Tr}\big((\log\rho_1)\rho_1\big) = S(\rho_1)\, .
\]
Since $S(\rho_1) \in \R$, Item~(\ref{item:se-sym}) yields $S(\Sigma_N) = \overline{S(T_N)} = S(\rho_1)$, establishing~\eqref{eq:se-isom}. Finally, in the time-homogeneous unitary setting of Definition~\ref{def:spacetime}, Theorem~\ref{prop:N-slices} gives $R_N = T_N$, whence $S(R_N) = S(\rho_1)$, which vanishes whenever $\rho_1$ is pure.
\eprf

\section{Concluding remarks} \label{sec:concl}
In this work, we have shown that three independently developed non-Hermitian generalizations of the density operator formalism into the temporal domain are mathematically equivalent up to Hermitian conjugation. For time-homogeneous unitary evolution, Theorem~\ref{prop:N-slices} establishes that the $N$-time quantum action of Diaz et al.~\cite{Diaz_2026a,Diaz_2026} factorizes as an ordered product of Jamio{\l}kowski operators, yielding the exact identity $R_N = T_N$ between $N$-time spacetime states and $N$-time KD states over time for every $N > 1$. For arbitrary sequential quantum channels, Theorem~\ref{thm:sdm-equiv} establishes that the $N$-time spacetime density matrix $\Sigma_N$ extending the constructions of Milekhin, Adamska, and Preskill~\cite{Milekhin_2025} and Guo~\cite{Guo_2025} satisfies $\Sigma_N = T_N^\dagger$, and hence $\Sigma_N = R_N^\dagger$ in the time-homogeneous unitary setting. Together with the basic properties collected in Proposition~\ref{prop:kd-properties}, these results establish a direct dictionary between quantum information theory, spacetime quantum mechanics, and timelike entanglement in quantum field theory. Moreover, expressing these operators as KD states over time yields a compact closed-form extension of both spacetime states and multi-time spacetime density matrices to arbitrary open-system dynamics. In light of these equivalences, the terms \emph{spacetime state} and \emph{spacetime density matrix} may be used interchangeably (up to Hermitian conjugation), just as \emph{quantum state} and \emph{density matrix} are used interchangeably in standard quantum theory. Rather than retaining the separate terminology \emph{KD state over time}, it is therefore natural to refer to the general construction simply as a spacetime state or spacetime density matrix, adding the prefix \emph{KD} whenever one wishes to emphasize its relationship to Kirkwood-Dirac correlators.

Building on the contraction bound $\|T_N\|_\infty \le 1$ established in Proposition~\ref{prop:kd-properties}~(\ref{item:kd-norm}), we have also addressed the problem of identifying a canonical entropy measure for spacetime states. Several non-Hermitian extensions of von~Neumann entropy have been studied in the literature and systematically compared in Ref.~\cite{Chen_2026}. Extending the logarithm by analytic continuation yields the pseudo-entropy~\cite{NTTTW21,DHMTT22}, which connects naturally to analytic continuations of spacelike R\'enyi entropies in conformal field theory and holography~\cite{Milekhin_2025,Guo_2025}, yet suffers from branch-cut ambiguities, potential divergences, and violations of standard entropy inequalities. Alternative proposals include the modified pseudo-entropy~\cite{Tu22,Salek_2014,Cipolloni_2023}, which replaces eigenvalues inside the logarithm by their moduli, and two singular-value-based measures---the SVD entropy~\cite{PaTK23} and the ABB entropy~\cite{Alter_2000,Chen_2026}---which evaluate the Shannon entropy of normalized singular values and normalized squared singular values, respectively, yielding real, non-negative quantities bounded by the logarithm of the Hilbert-space dimension. In contrast, extending the axiomatic characterization of PDM entropy~\cite{Fullwood_Yang_2026} to the space $\mathcal{C}^1$ of finite-dimensional unit-trace contractions, Theorem~\ref{thm:spacetime-entropy} shows that four simple requirements---reduction to von~Neumann entropy, continuity, strong additivity over the closed unit disk $\mathbb{D}$, and the polar expectation value property of Definition~\ref{def:mod-exp}---isolate a unique spacetime entropy functional $S(X) = -\mathrm{Tr}(X\log|X|) = -\mathrm{Tr}(X\log\sqrt{XX^\dagger})$. While agreeing with modified pseudo-entropy on all normal contractions (including PDMs), this spacetime entropy remains anchored in the functional calculus through the polar decomposition and avoids the branch-cut ambiguities of pseudo-entropy. Furthermore, as established in Proposition~\ref{prop:spacetime-entropy-properties}, it retains key structural and dynamical features of the von~Neumann entropy, including additivity over tensor products, a singular-value bound on its magnitude, exact invariance $S(T_N) = S(T_N^{\dag}) = S(\rho_1)$ under arbitrary $N$-time isometric and unitary evolutions, and clean behavior under Hermitian conjugation via $S(T_N^{\dag}) = \overline{S(T_N)}$.

Several compelling directions remain open for future investigation. Although Theorem~\ref{thm:spacetime-entropy} provides a rigorous axiomatic foundation for the spacetime entropy $S(X) = -\mathrm{Tr}(X\log|X|)$, its operational interpretation and detailed behavior across general $N$-time quantum dynamics have yet to be explored in depth. In particular, it would be valuable to investigate how $S(T_N)$ behaves under paradigmatic open-system channels and unitary evolutions, to clarify the physical significance of a nonzero imaginary part in $S(T_N)$ (which flips sign under the ordering reversal $T_N \mapsto \Sigma_N = T_N^\dagger$), and to explore whether spacetime entropy yields a meaningful notion of quantum mutual information across time. Moreover, the polar expectation value formulation $S(X) = \mathrm{Tr}(X K(X))$ with $K(X) = -\log|X|$ opens the door to defining relative information measures, such as a spacetime relative entropy comparing distinct spacetime states via their associated modulus operators. Finally, extending both the operator equivalences established here and the axiomatic characterization of spacetime entropy to non-Markovian dynamics with memory, infinite-dimensional systems, and relativistic quantum field theory may shed further light on the relationship between timelike entanglement, causal structure, and spacetime geometry.
\\

\noindent \emph{Acknowledgements.---} We thank Nahuel L. Diaz for fruitful discussion which motivated this work, which is supported by the National Natural Science Foundation of China under grant no.~W2632013, and by the Hainan Provincial Natural Science Foundation of China under grant no.~126MS0010.
\\

\noindent \emph{AI Usage.---} Google Gemini 4 Neon was used for generating the proofs of mathematical statements, literature search, for extracting and condensing some definitions from the literature, and as a writing assistant. We take full responsibility for the validity of all generated text. 

\appendix

\section{Proofs of preliminary lemmas}\label{app:lemmas}

In this appendix, we provide the proofs of Lemmas~\ref{lem:complex-strong-add} and~\ref{lem:unit-polygon}, which are used in the proof of Theorem~\ref{thm:spacetime-entropy}.

\bprf[Proof of Lemma~\ref{lem:complex-strong-add}]
Define $h^0 \colon \Q(\mathbb{D}) \to \C$ by $h^0(z_1,\ldots,z_n) = -\sum_{j=1}^{n} z_j \log|z_j|$. For any $p \in \Q_k(\mathbb{D})$ and $q^{(i)} \in \Q_{m_i}(\mathbb{D})$, using $\sum_{j=1}^{m_i} q_{ij} = 1$ and the convention $0\log 0 = 0$ yields
\begin{align*}
h^0\!\left(\bigoplus_{i=1}^{k} p_i q^{(i)}\right)
&= -\sum_{i=1}^{k} \sum_{j=1}^{m_i} p_i q_{ij}\log|p_i q_{ij}|
= -\sum_{i=1}^{k} p_i\log|p_i| \sum_{j=1}^{m_i} q_{ij} - \sum_{i=1}^{k} p_i \sum_{j=1}^{m_i} q_{ij}\log|q_{ij}| \\
&= h^0(p_1,\ldots,p_k) + \sum_{i=1}^{k} p_i h^0\big(q^{(i)}\big)\, .
\end{align*}
Hence $r = h - h^0 \colon \Q(\mathbb{D}) \to \C$ is permutation invariant, continuous on each $\Q_n(\mathbb{D})$, strongly additive on $\Q(\mathbb{D})$, and satisfies $r(p) = 0$ for all $p \in \Delta$. We show that $r \equiv 0$.

First, appending or removing a zero entry leaves $r$ unchanged. Indeed, for any $z \in \Q_n(\mathbb{D})$, strong additivity applied to $(z_1,\ldots,z_n,0) = 1\cdot z \oplus 0\cdot (1)$ with outer vector $(1,0) \in \Delta_2$ gives $r(z_1,\ldots,z_n,0) = r(1,0) + r(z) = r(z)$. Similarly, if $z \in \Q_n(\mathbb{D})$ has two entries $z_1 = a_1 u$ and $z_2 = a_2 u$ with $u \in \mathbb{T}$, $a_1, a_2 > 0$, and $a = a_1 + a_2 \le 1$, then writing $z = (au)(a_1/a, a_2/a) \oplus z_3(1) \oplus \cdots \oplus z_n(1)$ and using $(a_1/a, a_2/a) \in \Delta_2$ yields
\be\label{eq:merge-disk}
r(a_1 u, a_2 u, z_3,\ldots,z_n) = r(au, z_3,\ldots,z_n)\, .
\ee

We next extend $r$ to a function $\tilde{r} \colon \mathcal{V}(\C) \to \C$. Given $z = (z_1,\ldots,z_n) \in \mathcal{V}_n(\C)$, choose $\alpha \in (0,1]$ such that $|\alpha z_j| \le 1$ for every $j \in \{1,\ldots,n\}$, and set
\[
\tilde{r}(z) = \frac{1}{\alpha}\, r(\alpha z_1,\ldots,\alpha z_n, 1-\alpha)\, .
\]
To verify that $\tilde{r}(z)$ is independent of $\alpha$, suppose $0 < \beta < \alpha \le 1$ both satisfy $|\alpha z_j| \le 1$ for all $j$, and set $t = \beta/\alpha \in (0,1)$. Splitting the positive entry $1-\beta = t(1-\alpha) + (1-t)$ via~\eqref{eq:merge-disk} and applying strong additivity with outer vector $(t, 1-t) \in \Delta_2$ gives
\begin{align*}
r(\beta z_1,\ldots,\beta z_n, 1-\beta)
&= r\big(t(\alpha z_1,\ldots,\alpha z_n, 1-\alpha) \oplus (1-t)(1)\big) \\
&= r(t,1-t) + t\, r(\alpha z_1,\ldots,\alpha z_n, 1-\alpha) + (1-t)r(1) \\
&= \frac{\beta}{\alpha}\, r(\alpha z_1,\ldots,\alpha z_n, 1-\alpha)\, ,
\end{align*}
which confirms that $\tilde{r}$ is well-defined. Taking $\alpha = 1$ and discarding the resulting zero entry shows that $\tilde{r}$ agrees with $r$ on $\Q(\mathbb{D})$. Moreover, for any $z^{(0)} \in \mathcal{V}_n(\C)$, fixing $\alpha = (1 + \max_j |z_j^{(0)}|)^{-1}$ gives a fixed formula for $\tilde{r}$ on an open neighborhood of $z^{(0)}$, so $\tilde{r}$ is continuous on each $\mathcal{V}_n(\C)$ and remains permutation invariant. Also, for any $w \in \mathcal{V}(\C)$ and $\alpha \in (0,1)$, choosing $\beta \in (0,1)$ with $|\beta w_l| \le 1$ for all $l$ and merging $\beta(1-\alpha)$ with $1-\beta$ via~\eqref{eq:merge-disk} gives
\be\label{eq:r-scale}
\tilde{r}(\alpha w, 1-\alpha) = \frac{1}{\beta}\, r\big(\beta\alpha w, \beta(1-\alpha), 1-\beta\big) = \frac{1}{\beta}\, r(\beta\alpha w, 1-\beta\alpha) = \alpha\, \tilde{r}(w)\, .
\ee

We now verify that $\tilde{r}$ satisfies strong additivity on all of $\mathcal{V}(\C)$. First let $p \in \Q_k(\mathbb{D})$ and $q^{(i)} \in \mathcal{V}_{m_i}(\C)$ ($i=1,\ldots,k$), set $w = \bigoplus_{i=1}^{k} p_i q^{(i)}$, and choose $\alpha \in (0,1)$ small enough that $|\alpha q_{ij}^{(i)}| \le 1$ for all $i, j$ (whence also $|\alpha w_l| \le 1$ for all $l$). Define $y^{(i)} = (\alpha q^{(i)}, 1-\alpha) \in \Q_{m_i+1}(\mathbb{D})$, so that $r(y^{(i)}) = \alpha\tilde{r}(q^{(i)})$ and, up to permutation, $\bigoplus_{i=1}^{k} p_i y^{(i)}$ equals $u = \alpha w \oplus (1-\alpha)p \in \Q(\mathbb{D})$. Applying strong additivity of $r$ on $\Q(\mathbb{D})$ yields
\be\label{eq:ru-one}
r(u) = r\!\left(\bigoplus_{i=1}^{k} p_i y^{(i)}\right) = r(p) + \sum_{i=1}^{k} p_i r\big(y^{(i)}\big) = r(p) + \alpha \sum_{i=1}^{k} p_i \tilde{r}\big(q^{(i)}\big)\, .
\ee
On the other hand, consider $v = \frac{1}{2}u \oplus \frac{1}{2}(1) \in \Q(\mathbb{D})$. Strong additivity with outer vector $(\frac{1}{2},\frac{1}{2}) \in \Delta_2$ gives $r(v) = \frac{1}{2}r(u)$. Splitting the last entry $\frac{1}{2} = \frac{1-\alpha}{2} + \frac{\alpha}{2}$ via~\eqref{eq:merge-disk} and regrouping yields $\frac{1}{2}(\alpha w, 1-\alpha) \oplus \frac{1}{2}((1-\alpha)p, \alpha)$, where both inner vectors lie in $\Q(\mathbb{D})$, so
\[
\frac{1}{2} r(u) = r(v) = \frac{1}{2} r(\alpha w, 1-\alpha) + \frac{1}{2} r\big((1-\alpha)p, \alpha\big) = \frac{1}{2}\alpha \tilde{r}(w) + \frac{1}{2}(1-\alpha) r(p)\, .
\]
Multiplying by $2$, equating with~\eqref{eq:ru-one}, and dividing by $\alpha$ gives $\tilde{r}(w) = r(p) + \sum_{i=1}^{k} p_i \tilde{r}(q^{(i)})$. Now let $p \in \mathcal{V}_k(\C)$ be arbitrary and choose $\alpha \in (0,1)$ so that $p' = (\alpha p, 1-\alpha) \in \Q_{k+1}(\mathbb{D})$. Applying the identity just proved to outer vector $p'$ and inner vectors $q^{(1)},\ldots,q^{(k)}, (1)$ gives
\[
\tilde{r}\!\left(\alpha\bigoplus_{i=1}^{k} p_i q^{(i)}, 1-\alpha\right) = r(\alpha p, 1-\alpha) + \alpha \sum_{i=1}^{k} p_i \tilde{r}\big(q^{(i)}\big)\, .
\]
Invoking~\eqref{eq:r-scale} and dividing by $\alpha$ yields
\be\label{eq:sa-all-C}
\tilde{r}\!\left(\bigoplus_{i=1}^{k} p_i q^{(i)}\right) = \tilde{r}(p) + \sum_{i=1}^{k} p_i \tilde{r}\big(q^{(i)}\big)
\ee
for all $p \in \mathcal{V}_k(\C)$ and $q^{(i)} \in \mathcal{V}_{m_i}(\C)$. In particular, the same argument as for~\eqref{eq:merge-disk} now shows that $\tilde{r}$ is unchanged by merging or splitting any two entries of the same phase (without requiring their moduli to sum to at most $1$).

Next, define $g \colon \C \to \C$ by $g(u) = \tilde{r}(1, u, -u)$, noting that $g(0) = 0$ and $g(-u) = g(u)$ by permutation invariance. For any $u, v \in \C$, evaluate $\tilde{r}$ on the vector $(\frac{1}{2}, \frac{1}{2}, \frac{u}{2}, -\frac{u}{2}, \frac{uv}{2}, -\frac{uv}{2}) \in \mathcal{V}_6(\C)$ in two ways. Grouping as $\frac{1}{2}(1,u,-u) \oplus \frac{1}{2}(1,uv,-uv)$ gives $\frac{1}{2}g(u) + \frac{1}{2}g(uv)$. Alternatively, merging the two entries $\frac{1}{2}$ into $1$ and grouping $(1, \frac{u}{2}, -\frac{u}{2}, \frac{uv}{2}, -\frac{uv}{2})$ as $1\cdot(1) \oplus \frac{u}{2}(1,v,-v) \oplus (-\frac{u}{2})(1)$ gives $g(u/2) + \frac{u}{2}g(v)$. Hence
\[
\frac{1}{2}g(u) + \frac{1}{2}g(uv) = g(u/2) + \frac{u}{2}g(v) \qquad \forall u, v \in \C\, .
\]
Setting $v = 0$ gives $g(u/2) = \frac{1}{2}g(u)$, and substituting this back yields $g(uv) = u g(v)$ for all $u, v \in \C$. Setting $u = -1$ and $v = 1$ gives $g(-1) = -g(1)$, while $g(-1) = g(1)$, whence $g(1) = 0$ and therefore $g(u) = u g(1) = 0$ for all $u \in \C$.

Consequently, appending or removing any opposite pair $(u, -u)$ leaves $\tilde{r}$ unchanged, since for any $z \in \mathcal{V}(\C)$ we may write $(z, u, -u) = 1\cdot z \oplus u\cdot(1) \oplus (-u)\cdot(1)$ and obtain $\tilde{r}(z, u, -u) = g(u) + \tilde{r}(z) = \tilde{r}(z)$. It follows immediately that $\tilde{r}(z) = 0$ whenever every entry of $z \in \mathcal{V}(\C)$ lies in $\R \cup i\R$. Indeed, after discarding zeros and merging entries sharing the same phase in $\{1, -1, i, -i\}$, the sum condition $\sum_j z_j = 1$ yields $(1 + a_-, -a_-, ib, -ib)$ for some $a_-, b \ge 0$, and splitting $1 + a_-$ into $1$ and $a_-$ followed by removing the opposite pairs $(a_-, -a_-)$ and $(ib, -ib)$ gives $\tilde{r}(z) = \tilde{r}(1) = 0$.

Now define a continuous function $\psi \colon \C \to \C$ by $\psi(x + iy) = \tilde{r}(1, -x, -iy, x + iy)$ for $x, y \in \R$. By the preceding paragraph, $\psi(w) = 0$ for every $w \in \R \cup i\R$. We claim that
\be\label{eq:r-sum-psi}
\tilde{r}(z_1,\ldots,z_n) = \sum_{j=1}^{n} \psi(z_j) \qquad \forall (z_1,\ldots,z_n) \in \mathcal{V}_n(\C)\, .
\ee
For $n = 1$, $z = (1)$ and $\tilde{r}(1) = 0 = \psi(1)$. For $n \ge 2$, write $z_j = x_j + i y_j$ ($x_j, y_j \in \R$), so that $\sum_{j=1}^{n} x_j = 1$ and $\sum_{j=1}^{n} y_j = 0$, and consider
\[
V = \left(\frac{1}{n},\ldots,\frac{1}{n},\, -\frac{x_1}{n},\ldots,-\frac{x_n}{n},\, -\frac{i y_1}{n},\ldots,-\frac{i y_n}{n},\, \frac{z_1}{n},\ldots,\frac{z_n}{n}\right) \in \mathcal{V}_{4n}(\C)\, .
\]
Grouping $V$ as $\bigoplus_{j=1}^{n} \frac{1}{n}(1, -x_j, -i y_j, z_j)$ yields $\tilde{r}(V) = \frac{1}{n}\sum_{j=1}^{n} \psi(z_j)$. On the other hand, letting $b \in \mathcal{V}_{3n}(\C)$ consist of $n$ copies of $\frac{1}{n-1}$ together with $-\frac{x_1}{n-1},\ldots,-\frac{x_n}{n-1}$ and $-\frac{i y_1}{n-1},\ldots,-\frac{i y_n}{n-1}$, every entry of $b$ lies in $\R \cup i\R$, so $\tilde{r}(b) = 0$. Up to permutation, $V = \frac{1}{n} z \oplus (1 - \frac{1}{n}) b$, whence $\tilde{r}(V) = \frac{1}{n}\tilde{r}(z)$. Equating both expressions for $\tilde{r}(V)$ and multiplying by $n$ establishes~\eqref{eq:r-sum-psi}.

We now determine $\psi$. For any $w = x + iy \in \C$ and any $s, t > 0$, splitting $(s+t)w$ into $sw$ and $tw$ inside $(1, -(s+t)x, -i(s+t)y, (s+t)w)$ and applying~\eqref{eq:r-sum-psi} together with $\psi|_{\R \cup i\R} = 0$ gives $\psi((s+t)w) = \psi(sw) + \psi(tw)$. By induction this implies $\psi(qw) = q\psi(w)$ for every positive rational $q$, and continuity of $\psi$ extends this to $\psi(tw) = t\psi(w)$ for all $t > 0$. Moreover, $0 = \tilde{r}(1, w, -w) = \psi(w) + \psi(-w)$ gives $\psi(-w) = -\psi(w)$, so $\psi(tw) = t\psi(w)$ for all $t \in \R$ and $w \in \C$.

Define $\phi \colon \mathbb{T} \to \C$ by $\phi(u) = u^{-1}\psi(u)$, so that $\phi$ is continuous on $\mathbb{T}$ and $\psi(tu) = tu\phi(u)$ for all $t \in \R$ and $u \in \mathbb{T}$. Given $u = a + ib \in \mathbb{T}$ ($a, b \in \R$), apply~\eqref{eq:sa-all-C} with outer vector $(1, i, -i)$ and inner vectors $(1)$, $(1, -a, -ib, u)$, $(1)$. The resulting vector is $(1, i, -ia, b, iu, -i)$, whose entries other than $iu$ all lie in $\R \cup i\R$. Applying~\eqref{eq:r-sum-psi} gives $\psi(iu) = i\psi(u)$, and hence $\phi(iu) = \phi(u)$ for all $u \in \mathbb{T}$. Next, for any $u = a + ib \in \mathbb{T}$ and $v = c + id \in \mathbb{T}$ ($a, b, c, d \in \R$), apply~\eqref{eq:sa-all-C} with outer vector $(1, -a, -ib, u)$ and inner vectors $(1)$, $(1)$, $(1)$, $(1, -c, -id, v)$. The resulting vector is $(1, -a, -ib, u, -cu, -idu, uv)$, so~\eqref{eq:r-sum-psi} yields
\[
\psi(u) + \psi(-cu) + \psi(-idu) + \psi(uv) = \psi(u) + u\psi(v)\, .
\]
Cancelling $\psi(u)$ and substituting $\psi(-cu) = -cu\phi(u)$, $\psi(-idu) = -d\psi(iu) = -idu\phi(u)$, $\psi(uv) = uv\phi(uv)$, and $u\psi(v) = uv\phi(v)$ gives
\[
uv\big(\phi(uv) - \phi(u) - \phi(v)\big) = 0\, ,
\]
so $\phi(uv) = \phi(u) + \phi(v)$ for all $u, v \in \mathbb{T}$. For every root of unity $\zeta \in \mathbb{T}$ satisfying $\zeta^m = 1$ ($m \in \mathbb{N}$), it follows that $m\phi(\zeta) = \phi(\zeta^m) = \phi(1) = 0$, so $\phi(\zeta) = 0$. Since the roots of unity are dense in $\mathbb{T}$ and $\phi$ is continuous, $\phi \equiv 0$. Therefore $\psi \equiv 0$, which by~\eqref{eq:r-sum-psi} yields $\tilde{r} \equiv 0$, and hence $r \equiv 0$ on $\Q(\mathbb{D})$, establishing~\eqref{eq:h-complex-formula}.
\eprf

\bprf[Proof of Lemma~\ref{lem:unit-polygon}]
We proceed by induction on $m \ge 1$. For $m = 1$, the hypotheses $c_1 \in (0,1]$ and $c_1 \ge 1$ force $c_1 = 1$, so $z_1 = 1$ suffices. Now assume the claim holds for $m - 1$ ($m \ge 2$), and let $c_1,\ldots,c_m \in (0,1]$ satisfy $\sum_{t=1}^{m} c_t \ge 1$. Set $d = \min(1, c_{m-1} + c_m) \in (0,1]$. Since $c_{m-1}, c_m \in (0,1]$, we have $|c_{m-1} - c_m| < \max(c_{m-1}, c_m) \le 1$ and $|c_{m-1} - c_m| < c_{m-1} + c_m$, whence
\[
|c_{m-1} - c_m| < d \le c_{m-1} + c_m\, .
\]
As $\theta \mapsto |c_{m-1} + c_m e^{i\theta}|$ is continuous on $[0,\pi]$ and takes the values $c_{m-1} + c_m$ at $\theta = 0$ and $|c_{m-1} - c_m|$ at $\theta = \pi$, the intermediate value theorem yields $\theta \in [0,\pi)$ such that $|c_{m-1} + c_m e^{i\theta}| = d$. Writing $c_{m-1} + c_m e^{i\theta} = d u$ with $u \in \mathbb{T}$, observe that $c_1,\ldots,c_{m-2}, d$ lie in $(0,1]$ and satisfy $\sum_{t=1}^{m-2} c_t + d \ge 1$ (trivially when $d = 1$, and by hypothesis when $d = c_{m-1} + c_m$). By the induction hypothesis, there exist $z_1,\ldots,z_{m-2}, w \in \mathbb{T}$ such that $\sum_{t=1}^{m-2} c_t z_t + d w = 1$. Setting $z_{m-1} = w u^{-1} \in \mathbb{T}$ and $z_m = w u^{-1} e^{i\theta} \in \mathbb{T}$ yields $c_{m-1} z_{m-1} + c_m z_m = w u^{-1}(c_{m-1} + c_m e^{i\theta}) = d w$, and thus $\sum_{t=1}^{m} c_t z_t = 1$.
\eprf 

\addcontentsline{toc}{section}{\numberline{}Bibliography}
\bibliographystyle{quantum}
\bibliography{references}


\end{document}